\documentclass[draft]{agujournal2019}
\pdfoutput=1
\usepackage{url} 
\usepackage{bm}
\usepackage{lineno}
\usepackage[finalnew]{trackchanges} 
\usepackage{soul}
\usepackage{amsmath}

\journalname{JGR: Machine Learning and Computation}

\begin{document}

\title{A Virtual Solar Wind Monitor \add{at}\remove{for} Mars with Uncertainty Quantification using Gaussian Processes}

%
%

\authors{A. R. Azari\affil{1}, E. Abrahams\affil{2}, F. Sapienza\affil{2}, J. Halekas\affil{3}, J. Biersteker\affil{4}, D. L. Mitchell\affil{5}, F. Pérez\affil{2}, M. Marquette\affil{5}, M. J. Rutala\affil{6}, C. F. Bowers\affil{6}, C. M. Jackman\affil{6}, S. M. Curry\affil{7}}

\affiliation{1}{University of British Columbia, Vancouver, British Columbia, Canada}
\affiliation{2}{University of California, Berkeley, California, USA}
\affiliation{3}{University of Iowa, Iowa City, Iowa, USA}
\affiliation{4}{Massachusetts Institute of Technology, Cambridge, Massachusetts, USA}
\affiliation{5}{Space Sciences Laboratory, University of California, Berkeley, California, USA}
\affiliation{6}{Dublin Institute for Advanced Studies, Dunsink Observatory, Dublin 15, Ireland}
\affiliation{7}{University of Colorado, Boulder, Colorado, USA}

\correspondingauthor{A. R. Azari}{azari@eoas.ubc.ca}

\begin{keypoints}

\item We develop a continuous estimate of the solar wind upstream of Mars using Gaussian process regression on MAVEN data.
\item This model enables solar wind estimation at Mars with an R$^{2}$ of $\ge$ 0.95 for 66$\%$ of the time since late 2014.
\item This solar wind proxy is best used for robust statistical studies of Mars' space environment and the heliosphere. 
\end{keypoints}
%
%


\begin{abstract} 

Single spacecraft missions do not measure the pristine solar wind continuously because of the spacecrafts' orbital trajectory.
The infrequent \add{spatiotemporal} cadence of measurement fundamentally limits conclusions about solar wind-magnetosphere coupling throughout the solar system.
At Mars, such single spacecraft missions result in limitations for assessing the solar wind's role in causing lower altitude observations such as \remove{aurora}\add{auroral} dynamics or atmospheric loss. 
In this work, we detail the development of a virtual solar wind monitor from the Mars Atmosphere and Volatile Evolution (MAVEN) mission; a single spacecraft.
This virtual solar wind monitor \remove{is}\add{provides} a continuous estimate of the solar wind upstream from Mars with\remove{in} uncertainties\remove{on MAVEN data}.   
We specifically employ Gaussian process regression to estimate the upstream solar wind and \remove{error}\add{uncertainty} estimations that scale with the data sparsity of our real observations.
This proxy enables continuous solar wind estimation at Mars with representative uncertainties for the majority of the time since since late 2014.
We conclude \add{by discussing suggested uses}\remove{with usage guidelines} of this virtual solar wind monitor \remove{to enable subsequent}\add{for} statistical studies of the Mars space environment \add{and heliosphere}\remove{by the community}.

\end{abstract} 

\section*{Plain Language Summary} 

When a spacecraft orbits a planet, it travels through multiple spatial regions and it can be a long time between subsequent measurements of a region. 
This makes it difficult to understand how one region affects another as two regions are never measured at the same time. 
This is the scenario that the orbiting Mars Atmosphere and Volatile Evolution (MAVEN) spacecraft is in at Mars when measuring the solar wind. 
It is commonly accepted that the solar wind conditions, including magnetic field, velocity, and density, affects a planet's space environment.
However, because of MAVEN’s orbit, there is a large amount of uncertainty when estimating how the solar wind affect physical processes like atmospheric loss and auroral formation. 
In this work we create a continuous estimation, or virtual monitor, of the solar wind from MAVEN measurements. 
We do this by applying a machine learning method to estimate solar wind parameters and a predicted confidence, or \remove{error}\add{uncertainty}, in these estimates. 
These \remove{errors}\add{uncertainties} increase as MAVEN obtains less sampling and our confidence in the solar wind prediction decreases.
We conclude this work by sharing the suggested usage of this method in future studies of Mars \add{and the heliosphere}.

\section{Introduction}\label{intro}

The solar wind, a flow of plasma from the Sun, is a primary source of mass, energy,  and momentum in planetary space environments, including that of Mars.
At Mars the solar wind density, pressure, temperature, and magnetic field direction all influence the state of Mars solar wind interaction \cite<see discussion within>[]{Halekas2017a, Halekas2017b}.
This has broad reaching implications for various physical studies and human operations on Mars.
The solar wind impacts auroral emission \cite{Schneider2015, Chaffin2022, Haider2022, Atri2022, Girazian2022}, alters global scale magnetic field morphologies through draping and reconnection \cite{Brain2006a, Chai2019, Dubinin2019, DiBraccio2022, Azari2023, Bowers2023}, changes surface radiation \cite{Guo2017}, impacts space weather monitoring \cite{Lee2018, Lee2023, Green2023}, and affects our predictions and understanding of atmospheric loss \cite{Brain2016, Dong2018, Egan2019}. 
Mars' solar wind interaction is further complicated by the presence of intrinsic crustal fields \cite{Acuna1999, Connerney1999} which are spatially inhomogeneous \cite<see>[for recent models]{Langlais2019, Gao2021} and plays a combined role with the solar wind in driving physical phenomena at Mars. \add[added citations]{}
Unlike Earth \cite{King2005}, Mars has no continuous upstream solar wind monitor; thus measurements made in the near-Mars region lack context for any external driving \cite{Lee2023}.

The Mars Atmosphere and Volatile Evolution (MAVEN) spacecraft's orbital observations of the Mars space environment \cite{Jakosky2015} and solar wind instrumentation suite \cite{Halekas2015, Connerney2015, Mitchell2016} currently offers one of the most extensive historical and real-time measurements of the solar wind at Mars.
Because of MAVEN's orbit, however, estimations of the solar wind plasma moments and magnetic field are recorded at a non regular cadence.
\remove{While the shortest time between subsequent solar wind measurements averages 1.25 minutes, it is common given the orbit of MAVEN to observe multi-hour gaps multiple times a day.}
\add{While the shortest time between subsequent solar wind measurements is often seconds or minutes, it is common given the orbit of MAVEN to observe multi-hour gaps multiple times a day}.
Current proxy estimations of the solar wind are based on physical models \cite{Dewey2016, Keebler2022, Wang2023} or by estimating the relationship between the MAVEN measurements in the magnetosheath and the pristine solar wind through presumed dependency approximated by data science methods including artificial neural networks \cite{Ruhunusiri2018} or correlation analysis \cite{Hurley2018, Dong2019}. 
These methods result in proxies without real-time estimates of prediction uncertainties and are based on prior assumptions of the interplay between the magnetosheath and the upstream solar wind which may not be correct as shown in \citeA{Azari2023}.
These two limitations preclude the use of current proxies for large scale statistical studies with robust uncertainty quantification aimed at the discovery of new time-dependent physics at Mars. 
\add{One possible ideal solution, that is expected to become increasingly relevant in the near future with multi-spacecraft missions to Mars} \cite{Benkhoff2021, Sanchez-Cano2022, Lillis2022}\add{, is to integrate solar wind datasets from active missions such as Mars Express and Tianwen-1} \cite{Ramstad2017, Zou2021}\add{; but even this ideal case will still result in discontinuous estimates of the solar wind at Mars.}

The leading class of methods for accurate prediction of multi-dimensional data from previously obtained observations are machine learning methods. 
However, this flexible class of methods has not been used extensively in planetary science as compared to other scientific fields \cite{Azari2021}.
This is largely due to non-trivial challenges in applying, or in some cases, developing, these methods for physical knowledge gain. 
Particularly within Earth and planetary science, use of these methods for scientific discovery requires uncertainty quantification on sparse spatio-temporal data \cite<e.g. see discussions within>[]{Azari2020, Karpatne2019, Haynes2023, Poduval2023}.
For example, a primary goal of MAVEN aims to study atmospheric loss at Mars \cite{Jakosky2015} in which it's useful to estimate the upstream solar wind at the same time as estimating altitude dependent processes. 
\remove{But, by the nature of MAVEN's orbit, solar wind proxies for observations at low altitudes - a critical source of planetary neutrals - have larger, and currently unaccounted, uncertainties than when using a proxy at mid or upper altitudes.}
\add{But, by the nature of MAVEN's orbit, solar wind estimations at low altitudes have larger, and currently unaccounted, uncertainties than when using solar wind proxy estimations at mid or upper altitudes.}
\add{This is particularly troublesome  for studies that aim to identify and quantify altitude dependent processes and their relation to the solar wind and potential atmospheric loss} \cite<e.g.>{Jakosky2015, Jakosky2018, Girazian2019, Cangi2020, Ramstad2023, He2023}.
Including uncertainties that reflect the sparse sampling of the solar wind, a common challenge in machine learning applications, is critical when developing future proxies of the solar wind.

In this work we address these challenges by creating a virtual solar wind monitor \remove{for}\add{at} Mars.
This virtual monitor, or estimation of the solar wind's density, velocity, temperature, and magnetic field is based on data from the Solar Wind Ion Analyzer (SWIA) \cite{Halekas2015} and Magnetometer (MAG) \cite{Connerney2015} instruments on MAVEN. 
We employ Gaussian process regression, a machine learning method, to enable estimation of uncertainties that scale with the time since real measurement \cite<see>{Rasmussen2006}.
Our solar wind estimate results in larger \remove{errors}\add{uncertainties} at lower altitudes and enables robust uncertainty quantification for subsequent community use in inference and forecasting.

Our motivation in developing this virtual solar wind estimation is to enable statistical studies of the Mars space environment that need continuous solar wind information and estimations of uncertainty within solar wind approximation. 
For example, Gaussian processes were used to estimate the solar wind in \citeA{Azari2023}, enabling disambiguation of the crustal field from solar wind effects on the draped magnetic field morphology.
The developed virtual solar wind monitor can be used to advance understanding of the influence of the solar wind throughout Mars' space environment, from the drivers of atmospheric loss to characterization of space weather at Mars.  
It naturally incorporates the limitation of sparse sampling inherent to orbiting spacecraft measurements by scaling uncertainties with the lapsed time since last observation. 
Within the following manuscript, we detail the development of this virtual solar wind monitor for Mars along with its accuracy, and discuss how best to interpret and use the monitor and associated \remove{errors}\add{uncertainty predictions} for future scientific analyses.
\section{Methodology}\label{methods}
Within this section we detail the development of the Gaussian process based estimate of the upstream solar wind at Mars including: details on the original dataset used for training and testing (\ref{data}) and details on the machine learning model itself (\ref{ml}) and its implementation. 

\subsection{Original Data Source for Training and Assessment}\label{data}

MAVEN through the MAG and SWIA instruments measures the upstream solar wind density, velocity, temperature and magnetic field \cite{Connerney2015, Halekas2015, Halekas2017b}. 
The MAG suite is comprised of two fluxgate magnetometers that rapidly (32 Hz) measure the magnetic field as MAVEN orbits Mars \cite{Connerney2015}.
SWIA is a toroidal electrostatic analyzer and measures ions between 5 and 25,000 eV \cite{Halekas2015}. 
These two instrument datasets have been combined to produce upstream solar wind data sets at high cadence (300 measurements per orbit\add{, averaging 1.25 minutes between subsequent measurements}) of the pristine solar wind \cite{Halekas2017a} upstream of Mars' bow shock.
The parameters we utilize in this study from the combined driver data set include: the interplanetary magnetic field, IMF ($B_x$, $B_y$, $B_z$, $\lvert B \rvert $, [nT]), solar wind velocity ($v_x$, $v_y$, $v_z$, $\lvert v \rvert $, [km/s]), the proton temperature ($T_p$, [eV]) and the proton density ($n_p$, [per cc]). 
\remove{Each parameter is treated separately in our algorithm and it is worth noting this is true for the magnitude quantities and components as well.}
\add{Each parameter is fit with a separate model ($x$ time, $y$ parameter values) and it is worth noting this is true for the magnitude quantities and components as well.}
\add{This was done such that each parameter can be used for subsequent downstream parameter dependence analyses without a previous assumed dependence.}
The implications of this design choice are discussed further in Sections \ref{results} and \ref{conclusion}. 
All vector quantities are measured in the Mars-Solar Orbital frame in which $\hat X$ points from Mars to the Sun, $\hat Z$ points to Mars' ecliptic north, and $\hat Y$ completes the right-handed Cartesian set. 
We take this combined upstream driver data set of nine parameters as measured from the arrival of MAVEN to the Mars system (late 2014) up to the middle of 2023 for training, testing, and final prediction of the developed virtual solar wind monitor.
Data are normalized before usage for training, testing, and final prediction by removing the mean and subsequently scaled to the standard deviation range of the data for all outputs ($y$). 
Inputs (\remove{X}\add{$x$, time}) \remove{(or time)} are scaled linearly between 0 and 100.
These normalizations are reversed when providing final predictions.

\subsubsection{Development of a Test Set on Sparse Spatiotemporal Data}

An ideal test set, or dataset used to assess the performance of a predictive model, \add{should represent the performance seen by users of the model} \remove{should be representative of the overall performance of the model.}
This is challenging in the case of many sparse (rarely or unevenly sampled) spatiotemporal applications \cite<e.g.>{Karpatne2019, Azari2021} because \add{available training datasets are often sparser than the final predictions.}\remove{it is unknown if the available training dataset has a different distribution than the dataset that will be predicted to fill in the gaps.}  
\add{For example, consider a hypothetical case relevant to this work, in which a spacecraft measures the solar wind every minute for two hours, followed by a three hour gap, before returning to the solar wind for another two hours of minute resolution solar wind measurements.}
\add{Every data point in the original solar wind dataset will be at most be one minute from a true measurement, however if we use this dataset to create an hourly prediction, two predictions will be at least one hour from a true measurement.}
\add[v2]{This means that while 100$\%$ of the original dataset was within a minute of a real measurement, only 75$\%$ of an hourly cadence predicted dataset are within a minute of a real measurement (6 out of 8 points).} 
\remove{With this model, the primary factor influencing the accuracy of the proxy is the time lapsed since the most recent solar wind measurement.}
\add{Because the primary factor driving solar wind estimation accuracy is time to a real measurement, assessing model performance on test sets that randomly sample available training datasets will drastically over predict performance.}
\remove{In summary, this proxy will most often be used in places where the original dataset does not have data at all, which are inherently the most error prone regions of the proxy. 
If we use a test set randomly sampled from where MAVEN samples the solar wind, the performance will be largely overinflated when tested on other MAVEN samples as compared to its actual usage.}
\add{For this solar wind proxy we address this problem by generating a test set that has the same distribution as an hourly predicted dataset.}

\add{To develop a representative test set we resample, or remove, sections of the original data, to create a dataset that contains the same data gaps that an hourly estimate would contain.}
\remove{To develop a representative test set we resample the original data to create similar gaps, or time to real (spacecraft measured) solar wind measurement, to what a continuous ($~\sim$hourly) estimate would contain.}
\add{To do this we first estimate the  distribution of shortest time to a real measurement of an hourly prediction. This distribution is estimated with 14 bins set from 0 to 28 days.}
\add{We truncate this distribution at 28 days from a recent measurement since we expect, and show below, limited model accuracy estimating the solar wind more than 28 days from a measurement. Additionally predictions under 28 days contain the large majority, 95$\%$, of the total predicted dataset and the described binning approach for developing a test set is infeasible at increasing gaps.}
\add{Second, we set the number of samples that will be used to generate a new distribution. This number was limited to permit no less than 5 samples in each 2 day bin and results in 365 samples, or 0.5$\%$ of the original dataset.}
\add{Each sample is a continuous subset of the data consisting of 1,000 real measurements with one identified test point within that has the correct time to recent measurement (subsets are further discussed in Section }\ref{ml}).
\remove{More specifically, we implement 365 separate model runs to test the models performance that we use to assess 365 points that are distributed in the same sparse sampling as the final prediction.}
\add{Third, we iterate through the distribution identifying the correct number of samples from the original dataset to create a new distribution that represents the final predicted set. If possible, we use data points with the correct gap to a recent measurement. If the original dataset does not contain any samples with this sparsity (e.g. gaps over 12 days) we then randomly select a point from the original dataset and remove data to obtain the correct sparsity.}
\remove{We then randomly select 0.5$\%$ of this distribution as a test set, with each point representing one test point in the window that the model is run over.}

\begin{figure}[ht]
 \noindent\includegraphics[width=\textwidth]{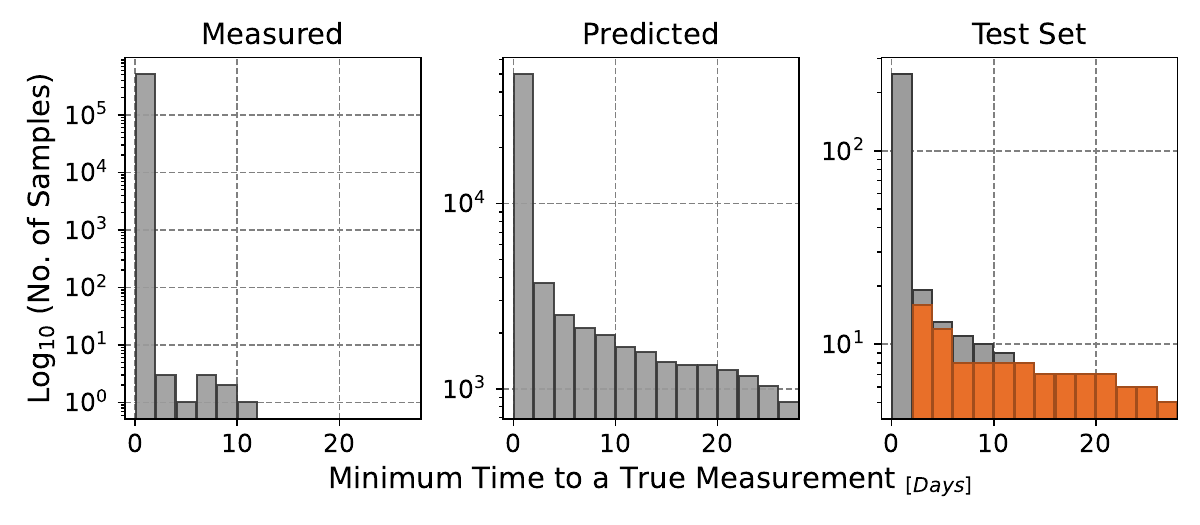}
 \caption{Development of test set that estimates the final proxy distribution. The left panel represents the measured solar wind from MAVEN MAG and SWIA instruments as distribution of time to closest measurement \cite{Connerney2015, Halekas2015, Halekas2017a}. The middle panel represents the distribution that would result from using a continuous proxy at an hour cadence. The right panel represents our new test set that is derived to match the distribution in the middle panel. Grey bars represent data that is sampled from the original (left) distribution and orange represents samples that are obtained by creating data gaps to the same frequency as the middle panel. The combined (grey and orange) represent the final test set.}
 \label{fig:testset}
 \end{figure}

In Figure \ref{fig:testset}, we demonstrate the original distribution of the solar wind data (left), the distribution that would result (middle) from sampling a continuous dataset every hour, and (right) our new test set distribution that we use to assess the algorithm. 
Because the algorithm performs best (discussed in following sections) at timestamps close to recent datapoints, we want the distribution in the right panel \add{(grey + orange)} to match as close as possible to that within the middle panel to accurately represent the final proxy's usage.
We use this test set to assess the predicted performance discussed in the following section.

\subsection{Solar Wind Estimation Model}\label{ml}

One goal in developing this monitor was to provide a physically uninformed,\add{ or} one \add{that is} not based on assumptions of the downstream draping interaction, estimation of the solar wind. 
While including physics knowledge often improves machine learning accuracy in planetary space physics \cite<e.g.>[]{Swiger2020, Azari2020} and more generally the inclusion of domain knowledge tends to improve model accuracy on natural science data \cite{Rudin2019, Karpatne2019}, the purpose of this monitor is to be subsequently used in other statistical studies to understand the Mars space environment and therefore we want to reduce any hypothesis bias that might be introduced by conforming to an ideal physical model. 
As a result, we specifically design this estimation of the solar wind to be purely data-driven, in order \remove{not to}\add{to not} introduce correlations with other physical variables future studies might wish to investigate. 
Our second goal was to provide robust estimations of uncertainty that reflect\remove{s} the sparse sampling that is inherent in spacecraft observations. 
%
%
By nature of the MAVEN orbit, there is no ground truth, or measurement, on what the solar wind should be when it is not directly measured \cite<see discussion on contemporaneous measurement challenges in>{Hurley2018}.
Creation of the test set as described above is one solution that can be used to approximate our understanding of the performance of the model in general, but this does not capture a direct estimate of the uncertainty for each individual measurement.
%
%
Until a real upstream solar wind monitor is available, we suggest that the best proxy for statistical studies is an estimate based on the nearest data available of a direct solar wind measurement with realistic uncertainties. 
Because of these two goals, we chose a machine learning model that specifically enables uncertainty quantification that is proportional to the data sampling and is designed around the covariance between data points.
\add{As a general rule of thumb, predicted uncertainties will be larger during measurement sparse, or highly variable, time periods.}

\subsubsection{Model Description: Gaussian Process Regression}

Gaussian process regression is a common method used for interpolating sparse geophysical datasets \cite<see usage suggestions in>{Tazi2023}.
In geostatistics Gaussian process models share historical legacy with kriging methods but subsequent mathematical generalization has redefined these methods as Gaussian process regression \cite<see>[for an introduction to Gaussian processes]{Rasmussen2006, Mackay1988}. 
This method is non-parametric, \remove{meaning that}\add{such that} this model is not limited to a particular linear combination\remove{s}.
This model differs from other machine learning methods by modeling trends between multidimensional inputs and outputs as a distribution over many possible functions. 
\remove{This framework of distributions over functions is what allows Gaussian processes to incorporate model uncertainty into posterior estimation.}

In \remove{the following}\add{this} model description we provide a high level overview of Gaussian processes\remove{before intersection back to our specific implementation}. 
Much of the following is based on more extensive derivations and discussions of implementations of Gaussian processes including within: \citeA{Rasmussen2006}, \citeA{Duvenaud2014} and \citeA{Tazi2023}. 
We follow the traditional mathematical formulation for a Gaussian process where we state that our model can be described as a deterministic function ($f$) in addition to a \add{stochastic data uncertainty}\remove{noise} term ($\epsilon$) that is normally distributed with a variance of $\sigma^{2}$ for some inputs, $x$ (Equation \ref{eqn:gp1}). \add{Note in our model this term does not include possible systematic measurement uncertainty.} We further describe $f(x)$ as a Gaussian process in Equation \ref{eqn:gp2}. \add{A Gaussian process is a collection of random variables, any finite number of which have a joint multivariate normal (Gaussian) distribution; in this case the random variables are the value of the function, $f(x)$ and $x$} \cite{Rasmussen2006}. 
\add{This is what allows for reasoning about a distribution of possible functions that can describe our observations.}

\add{It is worth noting one major implication of this formulation is that this model includes estimations of both stochastic data uncertainty, also known as aleatoric uncertainty} (see Equation \ref{eqn:gp1}), \add{and the model uncertainty also known as epistemic uncertainty which is estimated from distributions of possible functions} (see Equation \ref{eqn:gp2}). \add{Any systematic measurement errors, for example from the instrumentation itself, are not included in this model.} \add{It would be possible in future iterations of this model to include non random measurement errors a priori.}

\begin{linenomath*}
\begin{eqnarray} \label{eqn:gp1}
 y = f(x) + \epsilon, \;\;\;\; \epsilon \sim \mathcal{N}(0, \sigma^{2})
\end{eqnarray}
\end{linenomath*}

\begin{linenomath*}
\begin{eqnarray} \label{eqn:gp2}
 f(x) \sim \mathcal{GP}(m(x), k(x, x'))
\end{eqnarray}
\end{linenomath*}

In this formulation, $\mathcal{GP}$ denotes the Gaussian process model which describes the distribution of $f(x)$, $m(x)$ is the mean function which represents the mean of the process at $x$. 
\add{In our application as discussed above, $x$ is time and $y$ each parameter value, both normalized with each $y$ modeled separately, of the solar wind.}
Often, as \remove{in}\add{is} true in our implementation, we take $m(x)$ to be zero \add{everywhere}.  
$k(x, x')$ is the covariance function between data points, also known as the kernel. 
There are many well-studied kernel formulations, and a kernel should be picked to reflect how the response values at different inputs are expected to correlate \cite{Duvenaud2014}. 
%
%
In other words, implementing Gaussian processes requires the prior definition of how the user assumes data to be related to itself, or data covariance. 
We have chosen a rational quadratic covariance function, because this kernel represents data distributions that vary on multiple length scales.
This choice reflects the underlying physics of the solar wind that has multiple characteristic length scales and periodicities, varying on both short and long time scales \cite<e.g.>{Liu2021}.
There are limitations to this assumption's validity, as it is well known that certain observed solar wind properties have a periodic time period roughly relating to various solar (e.g. synodic Carrington rotation) periods \cite<e.g.>{Lee2017} which can affect the Mars space environment \cite<e.g.>{Liu2021}.
These are not accounted for in our model, and future efforts would be greatly improved by integrating this data-driven work and various physical solar wind propagation models \cite{Dewey2016, Barnard2022, Keebler2022, Wang2023}.

The kernel we use, \add{(rational quadratic)}, is defined mathematically below in equation \ref{eqn:model} from \citeA{Rasmussen2006}.

\begin{linenomath*}
\begin{eqnarray} \label{eqn:model}
 k(x,x^\prime) = \sigma^2 \left( 1 + \frac{\lVert x-x^\prime \rVert^2}{2 \alpha l^2} \right)^{-\alpha}   \nonumber \\
\end{eqnarray}
\end{linenomath*}

In this equation, the separation between $x$ and $x^\prime$ is the Euclidean distance between the feature vectors of two datapoints, $\sigma^2$ is the variance of the Gaussian process which scales the amplitude of the covariance between points, $\alpha$ is the scale mixture which influences the model flexibility, and $l$ is the length scale of the functional variation.
We evaluated other kernels (and possible combinations of kernels) and found this kernel to be the most representative of our data (e.g. capturing underlying patterns) and with the best (e.g. non-spurious) behavior in locations with limited sampling.  

\subsubsection{Model Implementation}

Gaussian processes are generally considered to be computationally expensive as compared to other machine learning methods given that they scale $\mathcal{O}(N^3)$ where $N$ is the number of points \cite{vanderwilk2020}. 
There are several approaches to allow for a reasonable convergence time including the use of graphics processing unit (GPU) computing \cite{Matthews2017, vanderwilk2020, Gardner2018} to accelerate parallel processing, the use of approximation methods for large datasets \cite{Rasmussen2006}, or even more generally breaking a larger dataset into representative subparts of $m$ size such that $\mathcal{O} \left(\frac{N^3}{m^2}\right)$ \cite{Candela2005}. 
There is a trade off between long computation time and accuracy, in which the most accurate models require larger data subsets $m$ but take longer to converge. 
In general running a Gaussian process of over 10,000 points is considered a large scale computing effort but one of even just 1,000 points is still a sizeable computing demand \cite<see package documentation within>{Gardner2018}.

We address the computational challenge by subsetting our data into chunks of 1,000 points and then implement the subsequent calculations onto GPFlow, a GPU enabled Gaussian process software package \cite{vanderwilk2020, Matthews2017}. 
We train and evaluate our model on the cloud based JupyterHub (Jupyter Meets the Earth) \cite{Pérez2019}. 
For each subset we use the L-BFGS-B optimization algorithm to estimate our kernel parameters as implemented via GPFlow's use of the the SciPy implementation \cite{scipy, Morales2011, Zhu1997, Byrd1995}. 
The L-BFGS family of optimization approaches is particularly memory-efficient because it approximates necessary inverse matrices without storing them, making it a useful choice for Gaussian processes.
Each \remove{subsets'}\add{subset's} $\sigma^2$ is initialized to a value of 3 and $l$ is set to an initial value of the first 10th percentile of the non-zero distances between each data point in the data subset and bounded between the 0th percentile and the 50th percentile.
$\sigma^2$ and $l$ are initialized to these values \remove{as they are relatively close to the aggregated final optimization but more importantly}\add{as they} represent our initial assumptions on the length scales and inherent variance and lead to good (i.e. limiting the overfit) performance on the test set.
\add{We further explored small alterations on these initial values to assess other reasonable initialization values that would lead to increased performance. We found that the initial values discussed above do not limit the final model and largely avoid problems of}\remove{Small explorations on these bounds were assessed to limit the effects} poor model estimations, for example solution collapse (where the model predicts the mean value of the subset of the data) or ringing \cite<see>{Duvenaud2014}.
The bounds on $l$ can be effectively interpreted to bounding the final length scale between no less than the shortest distance between datapoints present in the dataset and the 50th percentile.

In cases where we qualitatively observe poor model behavior, the solar wind parameter will usually be highly variable but sparsely sampled, such that the estimated length scales diminish to non realistic values and the predicted mean solution trends toward the prior value.   
We have limited the effects of these edge cases, where the solution converges to very low length scales, by limiting the bounds of $l$ and by taking subsets described above. 
By taking subsets we ensure that if a solution does collapse, then the predicted value at the worst case will output the uninformative mean of that subset.

This means that the worst predictions will trend toward the mean of the true data subset.
This is generally good behavior but a non ideal scenario for parameters in which their mean value is less representative of their mode (or most common) value (e.g. $B_{x}$ and $B_{y}$ which are bimodal).
It is for this reason that $\lvert B \rvert$ and $\lvert v \rvert$ are included as variables to provide a comparison point for users to assess the total magnitude and assess if the constituent values are underestimated. 
\section{Results}\label{results}

\begin{sidewaysfigure}
 \noindent\includegraphics[width=\textwidth]{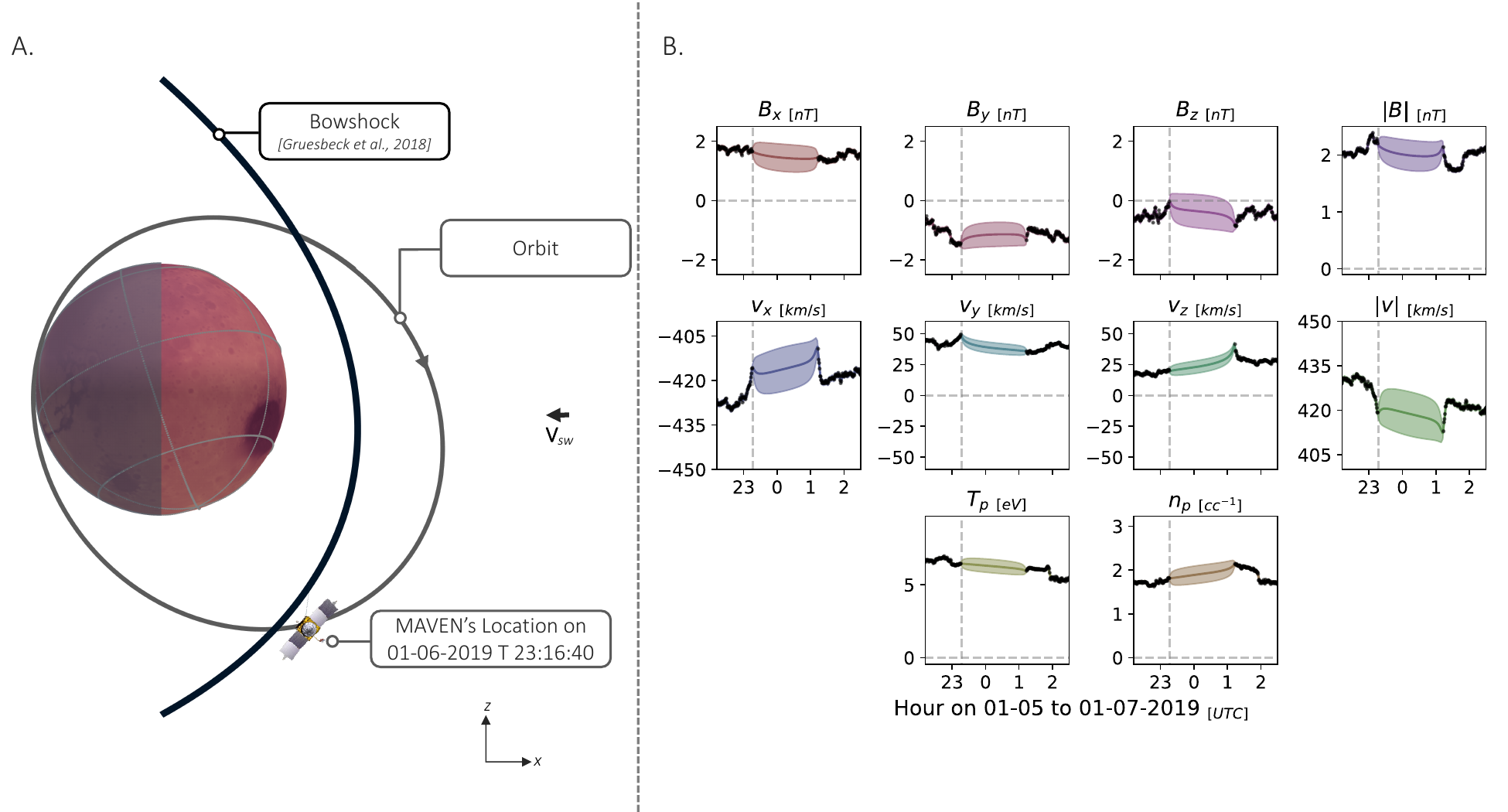}
 \caption{Example output of virtual solar wind monitor for MAVEN orbit number 8337. Panel A shows the MAVEN orbit along with the nominal bow shock location from \citeA{Gruesbeck2018}, the surface of Mars is elevation data from \citeA{Zuber1992} expected location on 01-06-2019T23:16:40 UTC as rotated with the NASA SPICE toolkit \cite{Acton2018, Acton1996}. Panel B shows training data (black dots), the mean prediction (colored line), and the standard deviation of this prediction (colored envelope). The vertical dashed line is where MAVEN's orbit passes inside the bow shock (location and time denoted in panel A) and stops measuring the solar wind.} 
 \label{fig:examplemonitor}
 \end{sidewaysfigure}

\subsection{Virtual Solar Wind Monitor Example from a Single Orbit}

Figure \ref{fig:examplemonitor} shows an example of the monitor results over a randomly selected orbit in 2019 (orbit number 8337). 
This is a common orientation of MAVEN as the spacecraft spends several hours inside the bow shock before traversing outside, to the solar wind. 
This figure demonstrates how when MAVEN no longer measures the solar wind after 01-06-2019T23:16:40 UTC, predicted \remove{errors}\add{uncertainties} (colored envelopes around the mean prediction) increase to reflect this transition.  
Meanwhile where MAVEN has measurements of the solar wind (black dots) the predicted \remove{errors}\add{uncertainties} are negligible given that the true value is known.
\add{Due to the rapid variability of the solar wind on the scale of minutes, predicted uncertainties often rise rapidly when a real measurement is not present. This can be interpreted to some extent as there being many possible functions that vary on short time scales which could represent the missing data.}

This figure also shows how, in the case that subsequent orbits measure similar values, the predicted mean value doesn't vary greatly (e.g. V$_{y}$) and if these values vary more within \remove{this}\add{a subset} period (e.g. V$_{x}$) then the predicted uncertainties will be larger.

\subsection{Evaluation Against a Spatially Representative Test Set}

\begin{sidewaysfigure}
 \noindent\includegraphics[width=0.85\textwidth]{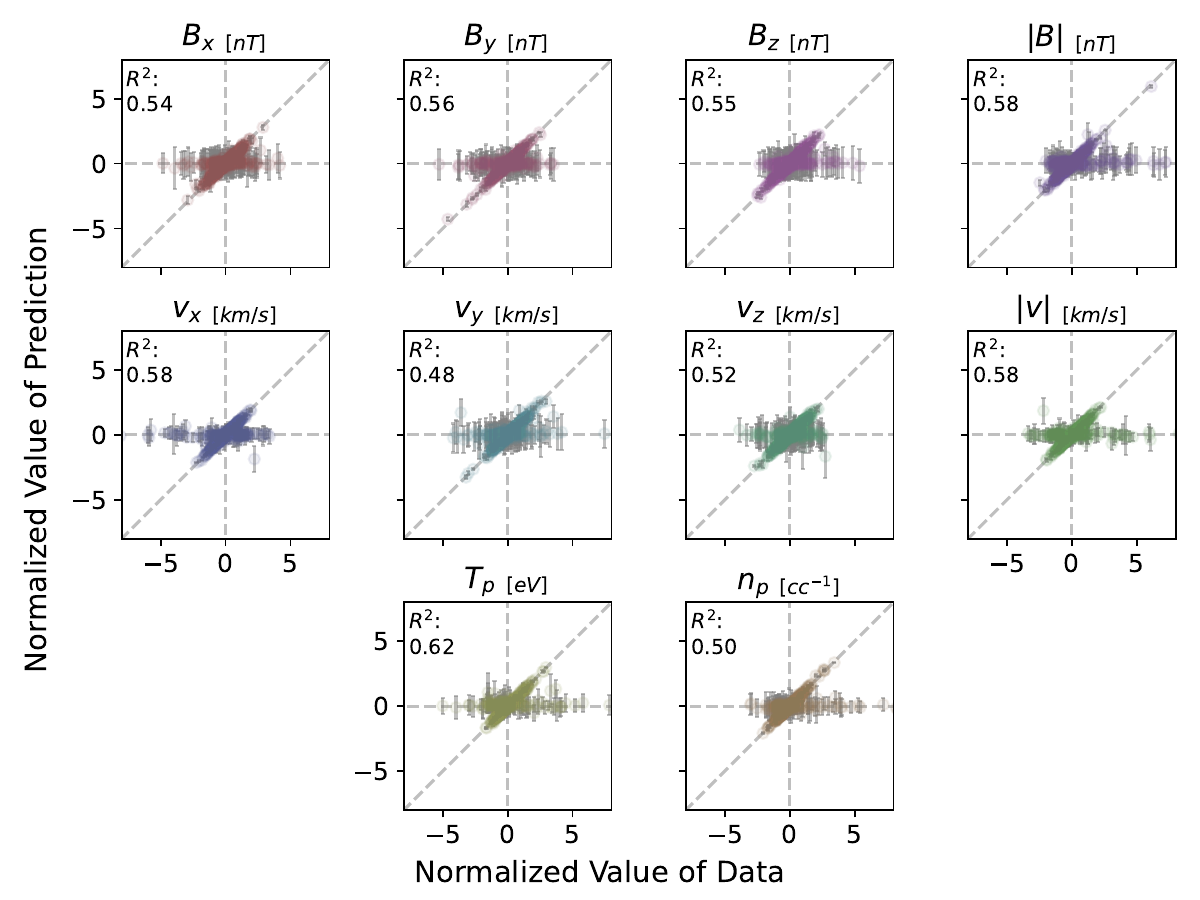}
 \caption{Performance of solar wind parameter estimation on test set as estimated by the coefficient of determination. Each plot includes the mean predicted value and predicted standard deviation in normalized space as compared to the true (test) data. The zero line in normalized space roughly maps to the median of the subset of data the Gaussian process model is evaluated over. The coefficient of determination, R$^{2}$, is provided as calculated on the un-normalized dataset to give an estimate of the true performance of the algorithm \cite{scikit-learn}.} 
 \label{fig:r2}
 \end{sidewaysfigure}

In Figure \ref{fig:r2} we compare the predicted mean values from our model to the true data values on our representative test set. 
This figure shows this comparison on the normalized values (in order to assess when the model is trending toward the mean of the dataset, zero in these plots). 
The coefficient of determination (R$^{2}$) is included in the top left corner calculated on the un-normalized values (the final model output).
A good R$^{2}$ (close to 1) generally indicates that our model captures the underlying variability of the data \cite<see>[for discussion of model data comparison metrics]{Liemohn2021}.
Overall, each solar wind parameter is well captured with an R$^{2}$ for all solar wind parameters of $\sim$ 0.5 or higher.
These R$^{2}$'s includes data up to 28 days from a true measurement. 
Within these plots the main source of model-data disagreement is when a predicted data point is very far from a true measurement, or in other words there is a large gap in true measurements.
For example, test points that are sufficiently far from a real measurement more frequently result in 0 (in normalized space, or the mean of the subset when un-normalized) on these figures.
Effectively this can be interpreted as the distance from a true measurement is sufficiently large enough that the solution trends to the preset of our model fit, the mean of the subset, and predicted \remove{errors}\add{uncertainties} rise accordingly.
This tends to lower the overall performance and can be seen in this figure as the points scattered around the zero line in normalized space.

\begin{figure}
 \noindent\includegraphics[width=0.85\textwidth]{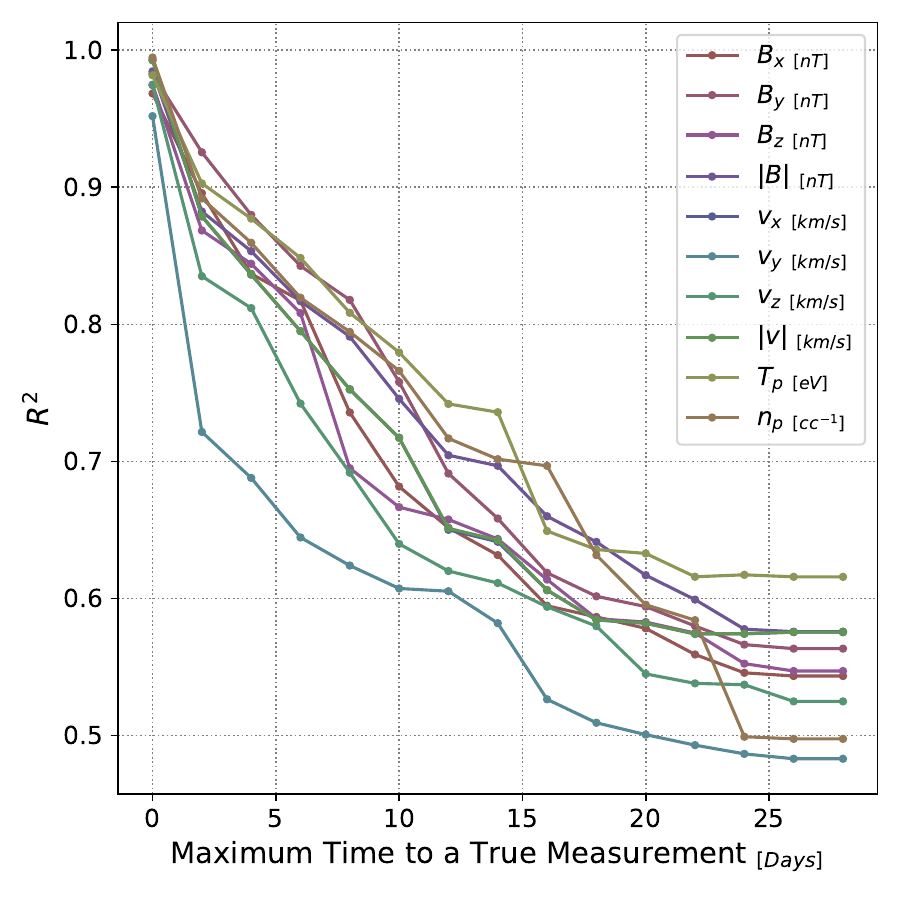}
 \caption{The coefficient of determination estimated as a function of the time to a real measurement of the solar wind. Points on the left of this plot are taken within 2 days of a solar wind value whereas points on the right of the plot are within 28 days of a real measurement.} 
 \label{fig:decline}
 \end{figure}

In Figure \ref{fig:decline} we show this relationship between performance via R$^{2}$ as a function of the maximum closest time to a recent measurement.
Overall the solar wind parameters are well captured, especially when estimated within 2 days of a true measurement, in which at least 95\% of the underlying variability is captured by our model fit (R$^{2}$ $>$ 0.95). 
However when the maximum closest time to a measurement is 28 days, the R$^{2}$ reduces to $\sim$ 0.5.
From this figure it becomes evident that the largest source of \remove{errors}\add{uncertainties} in our \remove{model}\add{predictions} is due to the time lapsed since a recent measurement. 
In general, these R$^{2}$ values indicate that \add{while} our model has generally captured the underlying variability of the data, it does not tell us if our predicted \remove{errors}\add{uncertainties} capture the expected values of the data\add{.}
In the following section we discuss our assessment of the predicted standard deviation values on the test set.

\subsection{Uncertainty Quantification Included as a Feature of the Proxy}

\begin{sidewaysfigure}
 \noindent\includegraphics[width=0.85\textwidth]{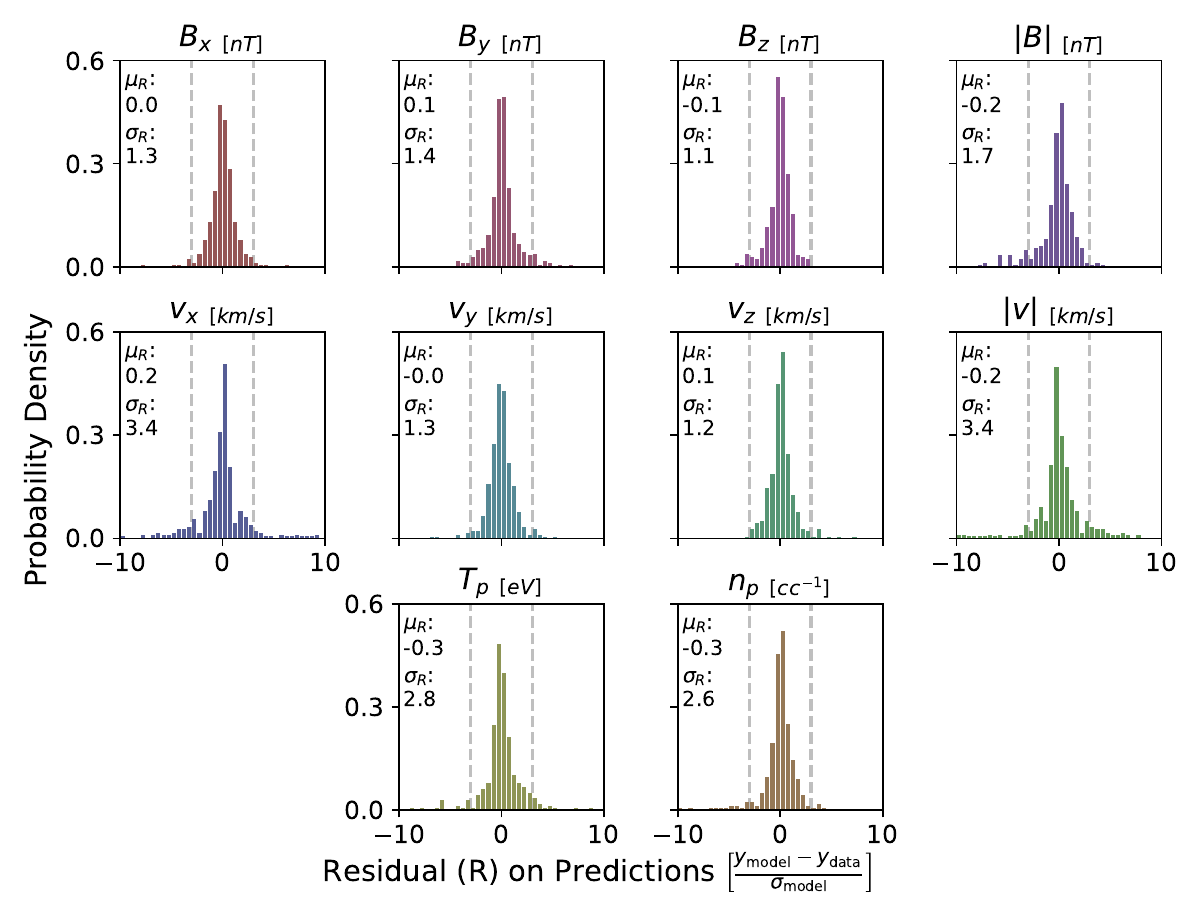}
 \caption{Estimate of the uncertainty quantification of the model as evaluated on the predicted standard deviations. Histograms are provided here of the prediction residual. The Fisher-Pearson coefficient of skewness (skew) is provided in the top left \cite{scipy, Zwillinger2000}. Vertical lines are included at the $\pm$ 3$R$ level.} 
 \label{fig:errordist}
 \end{sidewaysfigure}

Ideally our predicted standard deviations should capture the uncertainty between the model and the true data. 
In Figure \ref{fig:errordist} we plot the residuals of the model between the predicted value ($y_{model}$) and the observation, or true value ($y_{data}$), scaled to the predicted standard deviation ($\sigma_{model}$).
In the ideal case the mean of these residuals ($\mu_{R}$) should be near zero and the standard deviation $\sigma_{R}$ near one. 
Overall this is true of the mean of the residuals, with all features having very low bias toward over ($\mu_{R} > 0$) or under ($\mu_{R} < 0$) prediction.

However, $\sigma_{R}$ is feature dependent \remove{with }and generally trending larger than 1. 
For example you can interpret $B_{x}$, with a $\sigma_{R}$ of 1.3, to generally have uncertainty predictions \add{about 0.3 times smaller than a realistic representation.}\remove{on this are about 0.3 times too small to capture.} 
It is worth noting that several features have $\sigma_{R}$ above 2, in which the predicted uncertainties are underestimated significantly.
This is worth noting and while partially due to a mismatch between the model formulation and the dataset (see \ref{methods}) this is more due to the fact that these inaccuracies increase with the time lapse from a recent measurement.
Figure \ref{fig:errordist} represents the bulk aggregate over all the test set, including predictions close to a recent measurement (within hours) and predictions in which a relatively long time has passed (up to 28 days).

When using this proxy, it is worth knowing at what point the mean prediction (estimated via R$^{2}$) and the predicted uncertainties (estimated via $\sigma_{R}$) no longer capture the true solar wind.
In Table 1 we detail this by evaluating the assessment metrics for the entire test dataset divided into three ranges: all data within 2 days (66$\%$ of the time), data within 10 days of a solar wind measurement (80$\%$ of the time), and within 28 days (95$\%$ of the time). 
From this table it becomes clear that even up to 10 days from a true measurement our model fit provides both an accurate estimate of the true value of the solar wind and for most parameters, an accurate estimate of the uncertainties. 
And certain parameters (notably $B_{x}$, $B_{y}$, $B_{z}$, $V_{y}$, $V_{z}$) have reasonable predicted uncertainties, up to 28 days from a true measurement, effectively through the entire dataset.
This greatly expands the possible range of studies achievable for understanding the solar wind at Mars by extending estimates of the upstream solar wind accurately.

\subsection{Comparison to a Piecewise Linear Interpolation} 

\add{Table 1 additionally provides R$^{2}$ estimates on the test set for a piecewise linear interpolation.}
\add{From this comparison it is evident that the Gaussian process model provides similar accuracy to a linear interpolation at short time scales (within $\pm$ 2 days), but markedly improved performance far from recent measurements ($>$ 10 days).}
\add{Additionally, the Gaussian process based model provides reasonable uncertainty estimates far from recent measurements, which are not possible with a piecewise linear interpolation.}
\add{As a final note, it is possible to run a Gaussian process model with a linear kernel (or a combination of various kernels), but many of the advantages to such an approach for uncertainty quantification would be limited if approached in a piecewise manner.}
\add{We suggest that future solar wind proxy developments at Mars also compare performance against linear piecewise interpolations as a benchmarking exercise.}

\subsection{Suggested Use Cases and Limitations}

\add{This proxy, and the presented assessment statistics, are designed for large-scale statistical studies.}
\add{Expected typical use cases include assessing the influence of the solar wind parameters on the Mars space environment over multiple years with uncertainty, or providing an ongoing informed estimate of the solar wind at Mars for heliospheric or cross-mission studies.}
\add{It is not recommended that this version of the proxy is used to evaluate extreme events if MAVEN itself was not measuring the solar wind.}
\add{This is due to the smooth behavior of the estimates when MAVEN data is not present.}
\add{Similarly, the estimated R$^{2}$ performance, and uncertainty estimates, are assessed for cumulative use only.}
\add{This is important to note as the performance drops rapidly when real measurements are not present and often will trend to predicting a mean value of a data subset as designed.}
\add[v.2]{An example of using this proxy in a way that will have lower performance than reported would be to only use predictions that are two to four days from a recent measurement (e.g. not including estimates within $\pm$ two days).}
\add{While the reported range (see Table 1) of R$^{2}$ for predictions within $\pm$ two days is 0.95 to 0.99, the R$^{2}$ for predictions solely (e.g. not cumulatively including shorter gaps) from two to four days is at best 0.26 (for $B_x$), often trending to a mean value of the data subset}.
\add{This time period, two to four days from a recent measurement, only represents 3.3$\%$ of predictions and is significantly outweighed by the performance, and amount, of predictions that are close to a solar wind measurement.}
\add{This is the primary rationale to our suggestion to restrict usage of this version of the model to cumulative, rather than event based, studies.}
\add{To reason about this limitation, the provided model includes the time to a recent measurement and documentation on how to use these times.}
\add{We further direct interested users to the Open Research section to review the usage guidelines included in the model documentation.}

 \begin{table}
 \label{tab:stats}
 \caption{Estimated Model Performance for Maximum Times to a True Measurement and Comparison to Piecewise (PW) Linear Interpolation$^{a}$}
 \centering
 \begin{tabular}{l c c c}
 \hline
  Parameter & $R^2$, (PW $R^2$) & Mean of Residuals ($\mu_{R})$ & Stan. Dev. of Residuals ($\sigma_{R}$) \\
   & $\pm$ 2 days  & $\pm$ 2 days & $\pm$ 2 days \\
   & $\pm$ 10 days & $\pm$ 10 days & $\pm$ 10 days \\
   & $\pm$ 28 days & $\pm$ 28 days & $\pm$ 28 days \\
 \hline
   $B_{x}$ $_{[nT]}$ & 0.97 (0.98) & 0.0 & 1.0 \\
                     & 0.74 (0.72) & 0.0 & 1.1 \\
                     & 0.54 (0.44) & 0.0 & 1.3 \\
                     
   $B_{y}$ $_{[nT]}$ & 0.98 (0.98) & 0.1 & 1.0 \\ 
                     & 0.82 (0.77) & 0.1 & 1.2 \\
                     & 0.56 (0.48)  & 0.1 & 1.4 \\
                     
   $B_{z}$ $_{[nT]}$ & 0.97 (0.96) & 0.0 & 0.8 \\
                     & 0.70 (0.53) & 0.0 & 1.0  \\
                     & 0.55 (0.30) & -0.1 & 1.1\\
                     
   $\lvert B \rvert$ $_{[nT]}$ & 0.98 (0.98) & 0.0 & 1.0 \\
                               & 0.79 (0.76) & -0.1 & 1.3 \\
                               & 0.58 (0.55) & -0.2 & 1.7 \\
                               
   $v_{x}$ $_{[km/s]}$ & 0.99 (1.00)  & 0.1 & 1.0 \\
                       & 0.75 (0.68) & -0.1 & 2.2 \\
                       & 0.58 (0.49) & 0.2 & 3.4 \\
                       
   $v_{y}$ $_{[km/s]}$ & 0.95 (0.99) & -0.1 & 1.0 \\
                       & 0.62 (0.52) & 0.0 & 1.1 \\
                       & 0.48 (0.30) & 0.0 & 1.3 \\
                       
   $v_{z}$ $_{[km/s]}$ & 0.97 (0.96)  & 0.1 & 0.9 \\
                       & 0.69 (0.61) & 0.1 & 1.0 \\
                       & 0.52 (0.36) & 0.1 & 1.2 \\
                       
   $\lvert v \rvert$$_{[km/s]}$ & 0.99 (1.00) & -0.1 & 1.0 \\
                                & 0.75 (0.68)  & 0.0 & 2.2 \\
                                & 0.58 (0.49) & -0.2 & 3.4 \\
                                
   $T_{p}$ $_{[eV]}$ & 0.98 (0.99) & -0.1 & 0.9 \\
                     & 0.81 (0.64) &  0.0 & 1.7 \\
                     & 0.62 (0.43) & -0.3 & 2.8 \\
                     
   $n_{p}$ $_{[cc^{-1}]}$ & 0.99 (0.99)  & 0.0  & 0.8 \\
                          & 0.78 (0.69) & -0.3 & 2.4 \\
                          & 0.50 (0.42)  & -0.3 & 2.6 \\
 \hline
\multicolumn{4}{l}{$^{a}$Estimated on test set composed of 0.5$\%$ of resampled solar wind data from late 2014 to mid 2023.}
 \end{tabular}
 \end{table}

\section{Conclusion}\label{conclusion}
Estimating the solar wind to understand planetary-solar wind interactions from single spacecraft data has been a fundamental problem for the Mars community. 
Even with multi-spacecraft missions to Mars and other planets in the future \cite{Benkhoff2021, Lillis2022, Sanchez-Cano2022}, long duration concurrent measurements of the upstream solar wind are extremely unlikely.
In the absence of a real observational asset, we have implemented a machine learning method, Gaussian process regression, on MAVEN data to estimate the probable distributions of the solar wind, magnetic field (IMF), velocity, temperature, and density with both model and measurement uncertainties.
With this method we are able to gain highly accurate results of the solar wind within 2 days of a true MAVEN measurement (0.95 to 0.99 R$^2$, 66$\%$ of the dataset), reasonable estimations within 10 days (0.62 to 0.82 R$^2$, 80$\%$ of the dataset), and informed estimates within 28 days (0.48 to 0.62 R$^2$, 95$\%$ of the dataset).

Use of this proxy enables long-duration statistical studies that require robust uncertainty quantification \cite<e.g.>{Azari2023}.
However, due to the nature of the model employed, this proxy will be unable to capture any short or dynamic transitions of the solar wind, including transient events (e.g. coronal mass ejections).  
Thus, we recommend this model (and its associated predictions) is only used in concert with the predicted uncertainty estimations.

Subsequent improvements in this model are \remove{possible}\add{expected} through integrating this product with other missions' solar wind observations including Mars Express and Tianwen-1 \cite{Ramstad2017, Zou2021}.
Both missions have similar issues with discontinuous solar wind estimates \add{and combining these datasets will likely improve solar wind estimations.}
\add{These improvements}\remove{and} will require additional efforts in mission to mission inter-calibration before integration. 
Despite its limitations, the model described within enables highly accurate and continuous estimations of the solar wind for the majority of the ongoing\add{, and past, }lifetime of the MAVEN spacecraft.
We expect this model, and subsequent versions of this technique, to be valuable for future statistical studies of the Mars space environment, the heliosphere, and in estimating other planetary bodies' solar wind interactions. 

\section{Open Research}

The virtual solar wind monitor for Mars, a representative dataset, \add{usage examples}, and usage guidelines may be found at \url{https://github.com/abbyazari/vSWIM} \cite{Azari2024}. The MAVEN magnetic field data used in this study are available on the Planetary Data System at:
\url{https://pds-ppi.igpp.ucla.edu/mission/MAVEN} \cite{PDSMag}. The upstream driver file that combines SWIA and MAG data used to develop the virtual monitor and discussed within \cite{Halekas2017a} can be found online at \\
\noindent \url{https://homepage.physics.uiowa.edu/~jhalekas/drivers.html}. MAVEN data are also available at the MAVEN Science Data Center which can be accessed online at \\ \noindent \url{https://lasp.colorado.edu/maven/sdc/public/}. Results generated in this paper benefited from use of the Plotly, scipy, scikit-learn, GPFlow, and NASA SPICE software \cite{plotly, scipy, scikit-learn, vanderwilk2020, Matthews2017, Acton1996, Acton2018}. \add[note updated citation]{} Data management and workflows for this project benefited from the SQLite (sqlite.org) and the 2i2c projects (2i2c.org).


\acknowledgments
\add[v.2]{We thank our reviewers, including R. M. Dewey, for their attentive review which has positively impacted this work, such as their suggestion to compare this model to a piecewise linear interpolation.} This work was supported by the National Aeronautics and Space Administration (NASA) grant NNH10CC04C to the University of Colorado and by subcontract to Space Sciences Laboratory, University of California, Berkeley. The MAVEN project is supported by NASA through the Mars Exploration Program. We also acknowledge support from NASA's AI / ML Use Case Program, grant 80NSSC21K1370. This work was supported by the NSF Earth Cube Program under awards 1928406, 1928374. \add{ARA is supported by the Data Science Fellowship at the University of British Columbia through the Data Science Institute's Postdoctoral Matching Fund.} EA was supported by the National Science Foundation Graduate Research Fellowship under Grant No. DGE 1752814 and the Two Sigma PhD Fellowship. MJR is supported by Science Foundation Ireland Grant 18/FRL/6199, awarded to CMJ. CFB is supported by Irish Research Council Laureate Grant SOLMEX, awarded to CMJ.


\bibliography{refs}

\begin{thebibliography}{}

\bibitem [\protect \citeauthoryear {%
Acton%
}{%
Acton%
}{%
{\protect \APACyear {1996}}%
}]{%
Acton1996}
\APACinsertmetastar {%
Acton1996}%
\begin{APACrefauthors}%
Acton, C.%
\end{APACrefauthors}%
\unskip\
\newblock
\APACrefYearMonthDay{1996}{}{}.
\newblock
{\BBOQ}\APACrefatitle {Ancillary data services of {NASA's Navigation and
  Ancillary Information Facility}} {Ancillary data services of {NASA's
  Navigation and Ancillary Information Facility}}.{\BBCQ}
\newblock
\APACjournalVolNumPages{Planetary and Space Science}{44}{1}{65-70}.
\newblock
\begin{APACrefDOI} \doi{10.1016/0032-0633(95)00107-7} \end{APACrefDOI}
\PrintBackRefs{\CurrentBib}

\bibitem [\protect \citeauthoryear {%
Acton%
, Bachman%
, Semenov%
\BCBL {}\ \BBA {} Wright%
}{%
Acton%
\ \protect \BOthers {.}}{%
{\protect \APACyear {2018}}%
}]{%
Acton2018}
\APACinsertmetastar {%
Acton2018}%
\begin{APACrefauthors}%
Acton, C.%
, Bachman, N.%
, Semenov, B.%
\BCBL {}\ \BBA {} Wright, E.%
\end{APACrefauthors}%
\unskip\
\newblock
\APACrefYearMonthDay{2018}{}{}.
\newblock
{\BBOQ}\APACrefatitle {A look towards the future in the handling of space
  science mission geometry} {A look towards the future in the handling of space
  science mission geometry}.{\BBCQ}
\newblock
\APACjournalVolNumPages{Planetary and Space Science}{150}{}{9-12}.
\newblock
\begin{APACrefDOI} \doi{10.1016/j.pss.2017.02.013} \end{APACrefDOI}
\PrintBackRefs{\CurrentBib}

\bibitem [\protect \citeauthoryear {%
Acuña%
\ \protect \BOthers {.}}{%
Acuña%
\ \protect \BOthers {.}}{%
{\protect \APACyear {1999}}%
}]{%
Acuna1999}
\APACinsertmetastar {%
Acuna1999}%
\begin{APACrefauthors}%
Acuña, M\BPBI H.%
, Connerney, J\BPBI E\BPBI P.%
, Ness, N\BPBI F.%
, Lin, R\BPBI P.%
, Mitchell, D.%
, Carlson, C\BPBI W.%
\BDBL {}Cloutier, P.%
\end{APACrefauthors}%
\unskip\
\newblock
\APACrefYearMonthDay{1999}{}{}.
\newblock
{\BBOQ}\APACrefatitle {Global Distribution of Crustal Magnetization Discovered
  by the {Mars Global Surveyor MAG/ER Experiment}} {Global distribution of
  crustal magnetization discovered by the {Mars Global Surveyor MAG/ER
  Experiment}}.{\BBCQ}
\newblock
\APACjournalVolNumPages{Science}{284}{5415}{790-793}.
\newblock
\begin{APACrefDOI} \doi{10.1126/science.284.5415.790} \end{APACrefDOI}
\PrintBackRefs{\CurrentBib}

\bibitem [\protect \citeauthoryear {%
Atri%
, Dhuri%
, Simoni%
\BCBL {}\ \BBA {} Sreenivasan%
}{%
Atri%
\ \protect \BOthers {.}}{%
{\protect \APACyear {2022}}%
}]{%
Atri2022}
\APACinsertmetastar {%
Atri2022}%
\begin{APACrefauthors}%
Atri, D.%
, Dhuri, D.%
, Simoni, M.%
\BCBL {}\ \BBA {} Sreenivasan, K\BPBI R.%
\end{APACrefauthors}%
\unskip\
\newblock
\APACrefYearMonthDay{2022}{}{}.
\newblock
{\BBOQ}\APACrefatitle {Auroras on {Mars}: from discovery to new developments}
  {Auroras on {Mars}: from discovery to new developments}.{\BBCQ}
\newblock
\APACjournalVolNumPages{The European Physical Journal D}{76}{}{}.
\newblock
\begin{APACrefDOI} \doi{10.1140/epjd/s10053-022-00566-5} \end{APACrefDOI}
\PrintBackRefs{\CurrentBib}

\bibitem [\protect \citeauthoryear {%
Azari%
\ \protect \BOthers {.}}{%
Azari%
\ \protect \BOthers {.}}{%
{\protect \APACyear {2023}}%
}]{%
Azari2023}
\APACinsertmetastar {%
Azari2023}%
\begin{APACrefauthors}%
Azari, A\BPBI R.%
, Abrahams, E.%
, Sapienza, F.%
, Mitchell, D\BPBI L.%
, Biersteker, J.%
, Xu, S.%
\BDBL {}Dong, S., Y.and~Curry%
\end{APACrefauthors}%
\unskip\
\newblock
\APACrefYearMonthDay{2023}{}{}.
\newblock
{\BBOQ}\APACrefatitle {Magnetic Field Draping in Induced Magnetospheres:
  {Evidence} from the {MAVEN} Mission to {Mars}} {Magnetic field draping in
  induced magnetospheres: {Evidence} from the {MAVEN} mission to
  {Mars}}.{\BBCQ}
\newblock
\APACjournalVolNumPages{Journal of Geophysical Research: Space Physics}{}{}{}.
\newblock
\begin{APACrefDOI} \doi{10.1029/2023JA031546} \end{APACrefDOI}
\PrintBackRefs{\CurrentBib}

\bibitem [\protect \citeauthoryear {%
Azari%
\ \protect \BOthers {.}}{%
Azari%
\ \protect \BOthers {.}}{%
{\protect \APACyear {2021}}%
}]{%
Azari2021}
\APACinsertmetastar {%
Azari2021}%
\begin{APACrefauthors}%
Azari, A\BPBI R.%
, Biersteker, J\BPBI B.%
, Dewey, R\BPBI M.%
, Doran, G.%
, Forsberg, E\BPBI J.%
, Harris, C\BPBI D\BPBI K.%
\BDBL {}Ruhunusiri, S.%
\end{APACrefauthors}%
\unskip\
\newblock
\APACrefYearMonthDay{2021}{}{}.
\newblock
{\BBOQ}\APACrefatitle {Integrating Machine Learning for Planetary Science:
  {Perspectives} for the Next Decade} {Integrating machine learning for
  planetary science: {Perspectives} for the next decade}.{\BBCQ}
\newblock
\APACjournalVolNumPages{Bulletin of the AAS}{53}{4}{}.
\PrintBackRefs{\CurrentBib}

\bibitem [\protect \citeauthoryear {%
Azari%
\ \protect \BOthers {.}}{%
Azari%
\ \protect \BOthers {.}}{%
{\protect \APACyear {2024}}%
}]{%
Azari2024}
\APACinsertmetastar {%
Azari2024}%
\begin{APACrefauthors}%
Azari, A\BPBI R.%
, Halekas, J.%
, Hanley, K\BPBI G.%
, Abrahams, E.%
, Sapienza, F.%
, Biersteker, J.%
\BDBL {}Curry, S.%
\end{APACrefauthors}%
\unskip\
\newblock
\APACrefYearMonthDay{2024}{}{}.
\newblock
{\BBOQ}\APACrefatitle {{vSWIM}: v0.0.0} {{vSWIM}: v0.0.0}.{\BBCQ}
\newblock
\APACjournalVolNumPages{Zenodo}{}{}{}.
\newblock
\begin{APACrefURL} \url{https://doi.org/10.5281/zenodo.11106971}
  \end{APACrefURL}
\newblock
\begin{APACrefDOI} \doi{10.5281/zenodo.11106971} \end{APACrefDOI}
\PrintBackRefs{\CurrentBib}

\bibitem [\protect \citeauthoryear {%
Azari%
, Lockhart%
, Liemohn%
\BCBL {}\ \BBA {} Jia%
}{%
Azari%
\ \protect \BOthers {.}}{%
{\protect \APACyear {2020}}%
}]{%
Azari2020}
\APACinsertmetastar {%
Azari2020}%
\begin{APACrefauthors}%
Azari, A\BPBI R.%
, Lockhart, J\BPBI W.%
, Liemohn, M\BPBI W.%
\BCBL {}\ \BBA {} Jia, X.%
\end{APACrefauthors}%
\unskip\
\newblock
\APACrefYearMonthDay{2020}{}{}.
\newblock
{\BBOQ}\APACrefatitle {Incorporating Physical Knowledge Into Machine Learning
  for Planetary Space Physics} {Incorporating physical knowledge into machine
  learning for planetary space physics}.{\BBCQ}
\newblock
\APACjournalVolNumPages{Frontiers in Astronomy and Space Sciences}{7}{}{}.
\newblock
\begin{APACrefDOI} \doi{10.3389/fspas.2020.00036} \end{APACrefDOI}
\PrintBackRefs{\CurrentBib}

\bibitem [\protect \citeauthoryear {%
Barnard%
\ \BBA {} Owens%
}{%
Barnard%
\ \BBA {} Owens%
}{%
{\protect \APACyear {2022}}%
}]{%
Barnard2022}
\APACinsertmetastar {%
Barnard2022}%
\begin{APACrefauthors}%
Barnard, L.%
\BCBT {}\ \BBA {} Owens, M.%
\end{APACrefauthors}%
\unskip\
\newblock
\APACrefYearMonthDay{2022}{}{}.
\newblock
{\BBOQ}\APACrefatitle {{HUXt}—An open source, computationally efficient
  reduced-physics solar wind model, written in {P}ython} {{HUXt}—an open
  source, computationally efficient reduced-physics solar wind model, written
  in {P}ython}.{\BBCQ}
\newblock
\APACjournalVolNumPages{Frontiers in Physics}{10}{}{}.
\newblock
\begin{APACrefDOI} \doi{10.3389/fphy.2022.1005621} \end{APACrefDOI}
\PrintBackRefs{\CurrentBib}

\bibitem [\protect \citeauthoryear {%
Benkhoff%
\ \protect \BOthers {.}}{%
Benkhoff%
\ \protect \BOthers {.}}{%
{\protect \APACyear {2021}}%
}]{%
Benkhoff2021}
\APACinsertmetastar {%
Benkhoff2021}%
\begin{APACrefauthors}%
Benkhoff, J.%
, Murakami, G.%
, Baumjohann, W.%
, Besse, S.%
, Bunce, E.%
, Casale, M.%
\BDBL {}Zender, J.%
\end{APACrefauthors}%
\unskip\
\newblock
\APACrefYearMonthDay{2021}{}{}.
\newblock
{\BBOQ}\APACrefatitle {{BepiColombo} - Mission Overview and Science Goals}
  {{BepiColombo} - mission overview and science goals}.{\BBCQ}
\newblock
\APACjournalVolNumPages{Space Science Reviews}{217}{}{}.
\newblock
\begin{APACrefDOI} \doi{10.1007/s11214-021-00861-4} \end{APACrefDOI}
\PrintBackRefs{\CurrentBib}

\bibitem [\protect \citeauthoryear {%
Bowers%
\ \protect \BOthers {.}}{%
Bowers%
\ \protect \BOthers {.}}{%
{\protect \APACyear {2023}}%
}]{%
Bowers2023}
\APACinsertmetastar {%
Bowers2023}%
\begin{APACrefauthors}%
Bowers, C\BPBI F.%
, DiBraccio, G\BPBI A.%
, Slavin, J\BPBI A.%
, Gruesbeck, J\BPBI R.%
, Weber, T.%
, Xu, S.%
\BDBL {}Harada, Y.%
\end{APACrefauthors}%
\unskip\
\newblock
\APACrefYearMonthDay{2023}{}{}.
\newblock
{\BBOQ}\APACrefatitle {Exploring the Solar Wind-Planetary Interaction at
  {Mars}: {Implication} for Magnetic Reconnection} {Exploring the solar
  wind-planetary interaction at {Mars}: {Implication} for magnetic
  reconnection}.{\BBCQ}
\newblock
\APACjournalVolNumPages{Journal of Geophysical Research: Space
  Physics}{128}{2}{}.
\newblock
\begin{APACrefDOI} \doi{10.1029/2022JA030989} \end{APACrefDOI}
\PrintBackRefs{\CurrentBib}

\bibitem [\protect \citeauthoryear {%
Brain%
, Bagenal%
, Ma%
, Nilsson%
\BCBL {}\ \BBA {} Stenberg~Wieser%
}{%
Brain%
\ \protect \BOthers {.}}{%
{\protect \APACyear {2016}}%
}]{%
Brain2016}
\APACinsertmetastar {%
Brain2016}%
\begin{APACrefauthors}%
Brain, D\BPBI A.%
, Bagenal, F.%
, Ma, Y\BHBI J.%
, Nilsson, H.%
\BCBL {}\ \BBA {} Stenberg~Wieser, G.%
\end{APACrefauthors}%
\unskip\
\newblock
\APACrefYearMonthDay{2016}{}{}.
\newblock
{\BBOQ}\APACrefatitle {Atmospheric escape from unmagnetized bodies}
  {Atmospheric escape from unmagnetized bodies}.{\BBCQ}
\newblock
\APACjournalVolNumPages{Journal of Geophysical Research:
  Planets}{121}{12}{2364-2385}.
\newblock
\begin{APACrefDOI} \doi{10.1002/2016JE005162} \end{APACrefDOI}
\PrintBackRefs{\CurrentBib}

\bibitem [\protect \citeauthoryear {%
Brain%
, Mitchell%
\BCBL {}\ \BBA {} Halekas%
}{%
Brain%
\ \protect \BOthers {.}}{%
{\protect \APACyear {2006}}%
}]{%
Brain2006a}
\APACinsertmetastar {%
Brain2006a}%
\begin{APACrefauthors}%
Brain, D\BPBI A.%
, Mitchell, D\BPBI L.%
\BCBL {}\ \BBA {} Halekas, J\BPBI S.%
\end{APACrefauthors}%
\unskip\
\newblock
\APACrefYearMonthDay{2006}{}{}.
\newblock
{\BBOQ}\APACrefatitle {The magnetic field draping direction at {Mars} from
  {April} 1999 through {August} 2004} {The magnetic field draping direction at
  {Mars} from {April} 1999 through {August} 2004}.{\BBCQ}
\newblock
\APACjournalVolNumPages{Icarus}{182}{2}{464-473}.
\newblock
\begin{APACrefDOI} \doi{10.1016/j.icarus.2005.09.023} \end{APACrefDOI}
\PrintBackRefs{\CurrentBib}

\bibitem [\protect \citeauthoryear {%
Byrd%
, Lu%
, Nocedal%
\BCBL {}\ \BBA {} Zhu%
}{%
Byrd%
\ \protect \BOthers {.}}{%
{\protect \APACyear {1995}}%
}]{%
Byrd1995}
\APACinsertmetastar {%
Byrd1995}%
\begin{APACrefauthors}%
Byrd, R\BPBI H.%
, Lu, P.%
, Nocedal, J.%
\BCBL {}\ \BBA {} Zhu, C.%
\end{APACrefauthors}%
\unskip\
\newblock
\APACrefYearMonthDay{1995}{}{}.
\newblock
{\BBOQ}\APACrefatitle {A Limited Memory Algorithm for Bound Constrained
  Optimization} {A limited memory algorithm for bound constrained
  optimization}.{\BBCQ}
\newblock
\APACjournalVolNumPages{SIAM Journal on Scientific
  Computing}{16}{5}{1190-1208}.
\newblock
\begin{APACrefDOI} \doi{10.1137/0916069} \end{APACrefDOI}
\PrintBackRefs{\CurrentBib}

\bibitem [\protect \citeauthoryear {%
Cangi%
, Chaffin%
\BCBL {}\ \BBA {} Deighan%
}{%
Cangi%
\ \protect \BOthers {.}}{%
{\protect \APACyear {2020}}%
}]{%
Cangi2020}
\APACinsertmetastar {%
Cangi2020}%
\begin{APACrefauthors}%
Cangi, E\BPBI M.%
, Chaffin, M\BPBI S.%
\BCBL {}\ \BBA {} Deighan, J.%
\end{APACrefauthors}%
\unskip\
\newblock
\APACrefYearMonthDay{2020}{}{}.
\newblock
{\BBOQ}\APACrefatitle {Higher {M}artian Atmospheric Temperatures at All
  Altitudes Increase the {D/H} Fractionation Factor and Water Loss} {Higher
  {M}artian atmospheric temperatures at all altitudes increase the {D/H}
  fractionation factor and water loss}.{\BBCQ}
\newblock
\APACjournalVolNumPages{Journal of Geophysical Research:
  Planets}{125}{12}{e2020JE006626}.
\newblock
\begin{APACrefDOI} \doi{10.1029/2020JE006626} \end{APACrefDOI}
\PrintBackRefs{\CurrentBib}

\bibitem [\protect \citeauthoryear {%
Chaffin%
\ \protect \BOthers {.}}{%
Chaffin%
\ \protect \BOthers {.}}{%
{\protect \APACyear {2022}}%
}]{%
Chaffin2022}
\APACinsertmetastar {%
Chaffin2022}%
\begin{APACrefauthors}%
Chaffin, M\BPBI S.%
, Fowler, C\BPBI M.%
, Deighan, J.%
, Jain, S.%
, Holsclaw, G.%
, Hughes, A.%
\BDBL {}AlMatroushi, H.%
\end{APACrefauthors}%
\unskip\
\newblock
\APACrefYearMonthDay{2022}{}{}.
\newblock
{\BBOQ}\APACrefatitle {Patchy Proton Aurora at {Mars}: A Global View of Solar
  Wind Precipitation Across the {Martian} Dayside From {EMM/EMUS}} {Patchy
  proton aurora at {Mars}: A global view of solar wind precipitation across the
  {Martian} dayside from {EMM/EMUS}}.{\BBCQ}
\newblock
\APACjournalVolNumPages{Geophysical Research Letters}{49}{17}{}.
\newblock
\begin{APACrefDOI} \doi{10.1029/2022GL099881} \end{APACrefDOI}
\PrintBackRefs{\CurrentBib}

\bibitem [\protect \citeauthoryear {%
Chai%
\ \protect \BOthers {.}}{%
Chai%
\ \protect \BOthers {.}}{%
{\protect \APACyear {2019}}%
}]{%
Chai2019}
\APACinsertmetastar {%
Chai2019}%
\begin{APACrefauthors}%
Chai, L.%
, Wan, W.%
, Wei, Y.%
, Zhang, T.%
, Exner, W.%
, Fraenz, M.%
\BDBL {}Zhong, J.%
\end{APACrefauthors}%
\unskip\
\newblock
\APACrefYearMonthDay{2019}{}{}.
\newblock
{\BBOQ}\APACrefatitle {The Induced Global Looping Magnetic Field on {Mars}}
  {The induced global looping magnetic field on {Mars}}.{\BBCQ}
\newblock
\APACjournalVolNumPages{The Astrophysical Journal}{871}{2}{L27}.
\newblock
\begin{APACrefDOI} \doi{10.3847/2041-8213/aaff6e} \end{APACrefDOI}
\PrintBackRefs{\CurrentBib}

\bibitem [\protect \citeauthoryear {%
J\BPBI E.~Connerney%
\ \protect \BOthers {.}}{%
J\BPBI E.~Connerney%
\ \protect \BOthers {.}}{%
{\protect \APACyear {1999}}%
}]{%
Connerney1999}
\APACinsertmetastar {%
Connerney1999}%
\begin{APACrefauthors}%
Connerney, J\BPBI E.%
, Acuña, M\BPBI H.%
, Wasilewski, P\BPBI J.%
, F., N\BPBI N.%
, Reme, H.%
, C., M.%
\BDBL {}Cloutier, P\BPBI A.%
\end{APACrefauthors}%
\unskip\
\newblock
\APACrefYearMonthDay{1999}{}{}.
\newblock
{\BBOQ}\APACrefatitle {Magnetic lineations in the ancient crust of Mars}
  {Magnetic lineations in the ancient crust of mars}.{\BBCQ}
\newblock
\APACjournalVolNumPages{Science}{284}{5415}{794-798}.
\newblock
\begin{APACrefDOI} \doi{10.1126/science.284.5415.794} \end{APACrefDOI}
\PrintBackRefs{\CurrentBib}

\bibitem [\protect \citeauthoryear {%
J\BPBI E\BPBI P.~Connerney%
\ \protect \BOthers {.}}{%
J\BPBI E\BPBI P.~Connerney%
\ \protect \BOthers {.}}{%
{\protect \APACyear {2015}}%
}]{%
Connerney2015}
\APACinsertmetastar {%
Connerney2015}%
\begin{APACrefauthors}%
Connerney, J\BPBI E\BPBI P.%
, Espley, J.%
, Lawton, P.%
, Murphy, S.%
, Odom, J.%
, Oliversen, R.%
\BCBL {}\ \BBA {} Sheppard, D.%
\end{APACrefauthors}%
\unskip\
\newblock
\APACrefYearMonthDay{2015}{}{}.
\newblock
{\BBOQ}\APACrefatitle {The {MAVEN} Magnetic Field Investigation} {The {MAVEN}
  magnetic field investigation}.{\BBCQ}
\newblock
\APACjournalVolNumPages{Space Science Reviews}{195}{12}{257-291}.
\newblock
\begin{APACrefDOI} \doi{10.1007/s11214-015-0169-4} \end{APACrefDOI}
\PrintBackRefs{\CurrentBib}

\bibitem [\protect \citeauthoryear {%
Dewey%
\ \protect \BOthers {.}}{%
Dewey%
\ \protect \BOthers {.}}{%
{\protect \APACyear {2016}}%
}]{%
Dewey2016}
\APACinsertmetastar {%
Dewey2016}%
\begin{APACrefauthors}%
Dewey, R\BPBI M.%
, Baker, D\BPBI N.%
, Mays, M\BPBI L.%
, Brain, D\BPBI A.%
, Jakosky, B\BPBI M.%
, Halekas, J\BPBI S.%
\BDBL {}Lee, C\BPBI O.%
\end{APACrefauthors}%
\unskip\
\newblock
\APACrefYearMonthDay{2016}{}{}.
\newblock
{\BBOQ}\APACrefatitle {Continuous solar wind forcing knowledge: Providing
  continuous conditions at {Mars} with the {WSA-ENLIL+Cone} model} {Continuous
  solar wind forcing knowledge: Providing continuous conditions at {Mars} with
  the {WSA-ENLIL+Cone} model}.{\BBCQ}
\newblock
\APACjournalVolNumPages{Journal of Geophysical Research: Space
  Physics}{121}{7}{6207-6222}.
\newblock
\begin{APACrefDOI} \doi{10.1002/2015JA021941} \end{APACrefDOI}
\PrintBackRefs{\CurrentBib}

\bibitem [\protect \citeauthoryear {%
DiBraccio%
\ \protect \BOthers {.}}{%
DiBraccio%
\ \protect \BOthers {.}}{%
{\protect \APACyear {2022}}%
}]{%
DiBraccio2022}
\APACinsertmetastar {%
DiBraccio2022}%
\begin{APACrefauthors}%
DiBraccio, G\BPBI A.%
, Romanelli, N.%
, Bowers, C\BPBI F.%
, Gruesbeck, J\BPBI R.%
, Halekas, J\BPBI S.%
, Ruhunusiri, S.%
\BDBL {}Curry, S\BPBI M.%
\end{APACrefauthors}%
\unskip\
\newblock
\APACrefYearMonthDay{2022}{}{}.
\newblock
{\BBOQ}\APACrefatitle {A statistical investigation of factors influencing the
  magnetotail twist at {Mars}} {A statistical investigation of factors
  influencing the magnetotail twist at {Mars}}.{\BBCQ}
\newblock
\APACjournalVolNumPages{Geophysical Research Letters}{}{}{e2022GL098007}.
\newblock
\begin{APACrefDOI} \doi{10.1029/2022GL098007} \end{APACrefDOI}
\PrintBackRefs{\CurrentBib}

\bibitem [\protect \citeauthoryear {%
C.~Dong%
\ \protect \BOthers {.}}{%
C.~Dong%
\ \protect \BOthers {.}}{%
{\protect \APACyear {2018}}%
}]{%
Dong2018}
\APACinsertmetastar {%
Dong2018}%
\begin{APACrefauthors}%
Dong, C.%
, Lee, Y.%
, Ma, Y.%
, Lingam, M.%
, Bougher, S.%
, Luhmann, J.%
\BDBL {}Jakosky, B.%
\end{APACrefauthors}%
\unskip\
\newblock
\APACrefYearMonthDay{2018}{}{}.
\newblock
{\BBOQ}\APACrefatitle {Modeling Martian Atmospheric Losses over Time:
  Implications for Exoplanetary Climate Evolution and Habitability} {Modeling
  martian atmospheric losses over time: Implications for exoplanetary climate
  evolution and habitability}.{\BBCQ}
\newblock
\APACjournalVolNumPages{The Astrophysical Journal Letters}{859}{1}{L14}.
\newblock
\begin{APACrefDOI} \doi{10.3847/2041-8213/aac489} \end{APACrefDOI}
\PrintBackRefs{\CurrentBib}

\bibitem [\protect \citeauthoryear {%
Y.~Dong%
\ \protect \BOthers {.}}{%
Y.~Dong%
\ \protect \BOthers {.}}{%
{\protect \APACyear {2019}}%
}]{%
Dong2019}
\APACinsertmetastar {%
Dong2019}%
\begin{APACrefauthors}%
Dong, Y.%
, Fang, X.%
, Brain, D\BPBI A.%
, Hurley, D\BPBI M.%
, Halekas, J\BPBI S.%
, Espley, J\BPBI R.%
\BDBL {}Jakosky, B\BPBI M.%
\end{APACrefauthors}%
\unskip\
\newblock
\APACrefYearMonthDay{2019}{}{}.
\newblock
{\BBOQ}\APACrefatitle {Magnetic Field in the {Martian} Magnetosheath and the
  Application as an {IMF} Clock Angle Proxy} {Magnetic field in the {Martian}
  magnetosheath and the application as an {IMF} clock angle proxy}.{\BBCQ}
\newblock
\APACjournalVolNumPages{Journal of Geophysical Research: Space
  Physics}{124}{6}{4295-4313}.
\newblock
\begin{APACrefDOI} \doi{10.1029/2019JA026522} \end{APACrefDOI}
\PrintBackRefs{\CurrentBib}

\bibitem [\protect \citeauthoryear {%
Dubinin%
\ \protect \BOthers {.}}{%
Dubinin%
\ \protect \BOthers {.}}{%
{\protect \APACyear {2019}}%
}]{%
Dubinin2019}
\APACinsertmetastar {%
Dubinin2019}%
\begin{APACrefauthors}%
Dubinin, E.%
, Modolo, R.%
, Fraenz, M.%
, Päetzold, M.%
, Woch, J.%
, Chai, L.%
\BDBL {}Zelenyi, L.%
\end{APACrefauthors}%
\unskip\
\newblock
\APACrefYearMonthDay{2019}{}{}.
\newblock
{\BBOQ}\APACrefatitle {The Induced Magnetosphere of {Mars}: {Asymmetrical}
  Topology of the Magnetic Field Lines} {The induced magnetosphere of {Mars}:
  {Asymmetrical} topology of the magnetic field lines}.{\BBCQ}
\newblock
\APACjournalVolNumPages{Geophysical Research Letters}{46}{22}{12722-12730}.
\newblock
\begin{APACrefDOI} \doi{10.1029/2019GL084387} \end{APACrefDOI}
\PrintBackRefs{\CurrentBib}

\bibitem [\protect \citeauthoryear {%
Duvenaud%
}{%
Duvenaud%
}{%
{\protect \APACyear {2014}}%
}]{%
Duvenaud2014}
\APACinsertmetastar {%
Duvenaud2014}%
\begin{APACrefauthors}%
Duvenaud, D.%
\end{APACrefauthors}%
\unskip\
\newblock
\APACrefYearMonthDay{2014}{}{}.
\newblock
{\BBOQ}\APACrefatitle {Automatic model construction with {G}aussian processes}
  {Automatic model construction with {G}aussian processes}.{\BBCQ}
\newblock
\APACjournalVolNumPages{Apollo - University of Cambridge Repository}{}{}{}.
\newblock
\begin{APACrefDOI} \doi{10.17863/CAM.14087} \end{APACrefDOI}
\PrintBackRefs{\CurrentBib}

\bibitem [\protect \citeauthoryear {%
Egan%
, Jarvinen%
\BCBL {}\ \BBA {} Brain%
}{%
Egan%
\ \protect \BOthers {.}}{%
{\protect \APACyear {2019}}%
}]{%
Egan2019}
\APACinsertmetastar {%
Egan2019}%
\begin{APACrefauthors}%
Egan, H.%
, Jarvinen, R.%
\BCBL {}\ \BBA {} Brain, D.%
\end{APACrefauthors}%
\unskip\
\newblock
\APACrefYearMonthDay{2019}{}{}.
\newblock
{\BBOQ}\APACrefatitle {Stellar influence on heavy ion escape from unmagnetized
  exoplanets} {Stellar influence on heavy ion escape from unmagnetized
  exoplanets}.{\BBCQ}
\newblock
\APACjournalVolNumPages{Monthly Notices of the Royal Astronomical
  Society}{486}{1}{1283-1291}.
\newblock
\begin{APACrefDOI} \doi{10.1093/mnras/stz788} \end{APACrefDOI}
\PrintBackRefs{\CurrentBib}

\bibitem [\protect \citeauthoryear {%
Gao%
\ \protect \BOthers {.}}{%
Gao%
\ \protect \BOthers {.}}{%
{\protect \APACyear {2021}}%
}]{%
Gao2021}
\APACinsertmetastar {%
Gao2021}%
\begin{APACrefauthors}%
Gao, J\BPBI W.%
, Rong, Z\BPBI J.%
, Klinger, L.%
, Li, X\BPBI Z.%
, Liu, D.%
\BCBL {}\ \BBA {} Wei, Y.%
\end{APACrefauthors}%
\unskip\
\newblock
\APACrefYearMonthDay{2021}{}{}.
\newblock
{\BBOQ}\APACrefatitle {A Spherical Harmonic Martian Crustal Magnetic Field
  Model Combining Data Sets of MAVEN and MGS} {A spherical harmonic martian
  crustal magnetic field model combining data sets of maven and mgs}.{\BBCQ}
\newblock
\APACjournalVolNumPages{Earth and Space Science}{8}{10}{e2021EA001860}.
\newblock
\APACrefnote{e2021EA001860 2021EA001860}
\newblock
\begin{APACrefDOI} \doi{10.1029/2021EA001860} \end{APACrefDOI}
\PrintBackRefs{\CurrentBib}

\bibitem [\protect \citeauthoryear {%
Gardner%
, Pleiss%
, Bindel%
, Weinberger%
\BCBL {}\ \BBA {} Wilson%
}{%
Gardner%
\ \protect \BOthers {.}}{%
{\protect \APACyear {2018}}%
}]{%
Gardner2018}
\APACinsertmetastar {%
Gardner2018}%
\begin{APACrefauthors}%
Gardner, J\BPBI R.%
, Pleiss, G.%
, Bindel, D.%
, Weinberger, K\BPBI Q.%
\BCBL {}\ \BBA {} Wilson, A\BPBI G.%
\end{APACrefauthors}%
\unskip\
\newblock
\APACrefYearMonthDay{2018}{}{}.
\newblock
{\BBOQ}\APACrefatitle {GPyTorch: Blackbox Matrix-Matrix Gaussian Process
  Inference with GPU Acceleration} {Gpytorch: Blackbox matrix-matrix gaussian
  process inference with gpu acceleration}.{\BBCQ}
\newblock
\BIn{} \APACrefbtitle {Proceedings of the 32nd International Conference on
  Neural Information Processing Systems} {Proceedings of the 32nd international
  conference on neural information processing systems}\ (\BPG~7587–7597).
\newblock
\APACaddressPublisher{Red Hook, NY, USA}{Curran Associates Inc.}
\PrintBackRefs{\CurrentBib}

\bibitem [\protect \citeauthoryear {%
Girazian%
\ \protect \BOthers {.}}{%
Girazian%
\ \protect \BOthers {.}}{%
{\protect \APACyear {2019}}%
}]{%
Girazian2019}
\APACinsertmetastar {%
Girazian2019}%
\begin{APACrefauthors}%
Girazian, Z.%
, Halekas, J.%
, Morgan, D\BPBI D.%
, Kopf, A\BPBI J.%
, Gurnett, D\BPBI A.%
\BCBL {}\ \BBA {} Chu, F.%
\end{APACrefauthors}%
\unskip\
\newblock
\APACrefYearMonthDay{2019}{}{}.
\newblock
{\BBOQ}\APACrefatitle {The Effects of Solar Wind Dynamic Pressure on the
  Structure of the Topside Ionosphere of Mars} {The effects of solar wind
  dynamic pressure on the structure of the topside ionosphere of mars}.{\BBCQ}
\newblock
\APACjournalVolNumPages{Geophysical Research Letters}{46}{15}{8652-8662}.
\newblock
\begin{APACrefDOI} \doi{10.1029/2019GL083643} \end{APACrefDOI}
\PrintBackRefs{\CurrentBib}

\bibitem [\protect \citeauthoryear {%
Girazian%
\ \protect \BOthers {.}}{%
Girazian%
\ \protect \BOthers {.}}{%
{\protect \APACyear {2022}}%
}]{%
Girazian2022}
\APACinsertmetastar {%
Girazian2022}%
\begin{APACrefauthors}%
Girazian, Z.%
, Schneider, N\BPBI M.%
, Milby, Z.%
, Fang, X.%
, Halekas, J.%
, Weber, T.%
\BDBL {}Lee, C\BPBI O.%
\end{APACrefauthors}%
\unskip\
\newblock
\APACrefYearMonthDay{2022}{}{}.
\newblock
{\BBOQ}\APACrefatitle {Discrete Aurora at {Mars}: {Dependence} on Upstream
  Solar Wind Conditions} {Discrete aurora at {Mars}: {Dependence} on upstream
  solar wind conditions}.{\BBCQ}
\newblock
\APACjournalVolNumPages{Journal of Geophysical Research: Space
  Physics}{127}{4}{}.
\newblock
\begin{APACrefDOI} \doi{10.1029/2021JA030238} \end{APACrefDOI}
\PrintBackRefs{\CurrentBib}

\bibitem [\protect \citeauthoryear {%
Green%
, Dong%
, Hesse%
, Young%
\BCBL {}\ \BBA {} Airapetian%
}{%
Green%
\ \protect \BOthers {.}}{%
{\protect \APACyear {2022}}%
}]{%
Green2023}
\APACinsertmetastar {%
Green2023}%
\begin{APACrefauthors}%
Green, J\BPBI L.%
, Dong, C.%
, Hesse, M.%
, Young, C\BPBI A.%
\BCBL {}\ \BBA {} Airapetian, V.%
\end{APACrefauthors}%
\unskip\
\newblock
\APACrefYearMonthDay{2022}{}{}.
\newblock
{\BBOQ}\APACrefatitle {Space weather observations, modeling, and alerts in
  support of human exploration of {Mars}} {Space weather observations,
  modeling, and alerts in support of human exploration of {Mars}}.{\BBCQ}
\newblock
\APACjournalVolNumPages{Frontiers in Astronomy and Space Sciences}{9}{}{}.
\newblock
\begin{APACrefDOI} \doi{10.3389/fspas.2022.1023305} \end{APACrefDOI}
\PrintBackRefs{\CurrentBib}

\bibitem [\protect \citeauthoryear {%
Gruesbeck%
\ \protect \BOthers {.}}{%
Gruesbeck%
\ \protect \BOthers {.}}{%
{\protect \APACyear {2018}}%
}]{%
Gruesbeck2018}
\APACinsertmetastar {%
Gruesbeck2018}%
\begin{APACrefauthors}%
Gruesbeck, J\BPBI R.%
, Espley, J\BPBI R.%
, Connerney, J\BPBI E\BPBI P.%
, DiBraccio, G\BPBI A.%
, Soobiah, Y\BPBI I.%
, Brain, D.%
\BDBL {}Mitchell, D\BPBI L.%
\end{APACrefauthors}%
\unskip\
\newblock
\APACrefYearMonthDay{2018}{}{}.
\newblock
{\BBOQ}\APACrefatitle {The Three-Dimensional Bow Shock of Mars as Observed by
  MAVEN} {The three-dimensional bow shock of mars as observed by maven}.{\BBCQ}
\newblock
\APACjournalVolNumPages{Journal of Geophysical Research: Space
  Physics}{123}{6}{4542-4555}.
\newblock
\begin{APACrefDOI} \doi{10.1029/2018JA025366} \end{APACrefDOI}
\PrintBackRefs{\CurrentBib}

\bibitem [\protect \citeauthoryear {%
Guo%
\ \protect \BOthers {.}}{%
Guo%
\ \protect \BOthers {.}}{%
{\protect \APACyear {2017}}%
}]{%
Guo2017}
\APACinsertmetastar {%
Guo2017}%
\begin{APACrefauthors}%
Guo, J.%
, Slaba, T\BPBI C.%
, Zeitlin, C.%
, Wimmer-Schweingruber, R\BPBI F.%
, Badavi, F\BPBI F.%
, Böhm, E.%
\BDBL {}Rafkin, S.%
\end{APACrefauthors}%
\unskip\
\newblock
\APACrefYearMonthDay{2017}{}{}.
\newblock
{\BBOQ}\APACrefatitle {Dependence of the {Martian} radiation environment on
  atmospheric depth: {Modeling} and measurement} {Dependence of the {Martian}
  radiation environment on atmospheric depth: {Modeling} and
  measurement}.{\BBCQ}
\newblock
\APACjournalVolNumPages{Journal of Geophysical Research:
  Planets}{122}{2}{329-341}.
\newblock
\begin{APACrefDOI} \doi{10.1002/2016JE005206} \end{APACrefDOI}
\PrintBackRefs{\CurrentBib}

\bibitem [\protect \citeauthoryear {%
Haider%
\ \protect \BOthers {.}}{%
Haider%
\ \protect \BOthers {.}}{%
{\protect \APACyear {2022}}%
}]{%
Haider2022}
\APACinsertmetastar {%
Haider2022}%
\begin{APACrefauthors}%
Haider, S\BPBI A.%
, Mahajan, K\BPBI K.%
, Bougher, S\BPBI W.%
, Schneider, N\BPBI M.%
, Deighan, J.%
, Jain, S\BPBI K.%
\BCBL {}\ \BBA {} Gérard, J\BPBI C.%
\end{APACrefauthors}%
\unskip\
\newblock
\APACrefYearMonthDay{2022}{}{}.
\newblock
{\BBOQ}\APACrefatitle {Observations and Modeling of {Martian} Auroras}
  {Observations and modeling of {Martian} auroras}.{\BBCQ}
\newblock
\APACjournalVolNumPages{Space Science Reviews}{218}{}{}.
\newblock
\begin{APACrefDOI} \doi{10.1007/s11214-022-00906-2} \end{APACrefDOI}
\PrintBackRefs{\CurrentBib}

\bibitem [\protect \citeauthoryear {%
Halekas%
, Brain%
\BCBL {}\ \protect \BOthers {.}}{%
Halekas%
, Brain%
\BCBL {}\ \protect \BOthers {.}}{%
{\protect \APACyear {2017}}%
}]{%
Halekas2017b}
\APACinsertmetastar {%
Halekas2017b}%
\begin{APACrefauthors}%
Halekas, J\BPBI S.%
, Brain, D\BPBI A.%
, Luhmann, J\BPBI G.%
, DiBraccio, G\BPBI A.%
, Ruhunusiri, S.%
, Harada, Y.%
\BDBL {}Jakosky, B\BPBI M.%
\end{APACrefauthors}%
\unskip\
\newblock
\APACrefYearMonthDay{2017}{}{}.
\newblock
{\BBOQ}\APACrefatitle {Flows, Fields, and Forces in the Mars-Solar Wind
  Interaction} {Flows, fields, and forces in the mars-solar wind
  interaction}.{\BBCQ}
\newblock
\APACjournalVolNumPages{Journal of Geophysical Research: Space
  Physics}{122}{11}{}.
\newblock
\begin{APACrefDOI} \doi{10.1002/2017JA024772} \end{APACrefDOI}
\PrintBackRefs{\CurrentBib}

\bibitem [\protect \citeauthoryear {%
Halekas%
, Ruhunusiri%
\BCBL {}\ \protect \BOthers {.}}{%
Halekas%
, Ruhunusiri%
\BCBL {}\ \protect \BOthers {.}}{%
{\protect \APACyear {2017}}%
}]{%
Halekas2017a}
\APACinsertmetastar {%
Halekas2017a}%
\begin{APACrefauthors}%
Halekas, J\BPBI S.%
, Ruhunusiri, S.%
, Harada, Y.%
, Collinson, G.%
, Mitchell, D\BPBI L.%
, Mazelle, C.%
\BDBL {}Jakosky, B\BPBI M.%
\end{APACrefauthors}%
\unskip\
\newblock
\APACrefYearMonthDay{2017}{}{}.
\newblock
{\BBOQ}\APACrefatitle {Structure, dynamics, and seasonal variability of the
  {Mars}-solar wind interaction: {MAVEN Solar Wind Ion Analyzer} in-flight
  performance and science results} {Structure, dynamics, and seasonal
  variability of the {Mars}-solar wind interaction: {MAVEN Solar Wind Ion
  Analyzer} in-flight performance and science results}.{\BBCQ}
\newblock
\APACjournalVolNumPages{Journal of Geophysical Research: Space
  Physics}{122}{1}{547-578}.
\newblock
\begin{APACrefDOI} \doi{10.1002/2016JA023167} \end{APACrefDOI}
\PrintBackRefs{\CurrentBib}

\bibitem [\protect \citeauthoryear {%
Halekas%
\ \protect \BOthers {.}}{%
Halekas%
\ \protect \BOthers {.}}{%
{\protect \APACyear {2015}}%
}]{%
Halekas2015}
\APACinsertmetastar {%
Halekas2015}%
\begin{APACrefauthors}%
Halekas, J\BPBI S.%
, Taylor, E\BPBI R.%
, Dalton, G.%
, Johnson, G.%
, Curtis, D\BPBI W.%
, McFadden, J\BPBI P.%
\BDBL {}Jakosky, B\BPBI M.%
\end{APACrefauthors}%
\unskip\
\newblock
\APACrefYearMonthDay{2015}{}{}.
\newblock
{\BBOQ}\APACrefatitle {The Solar Wind Ion Analyzer for {MAVEN}} {The solar wind
  ion analyzer for {MAVEN}}.{\BBCQ}
\newblock
\APACjournalVolNumPages{Space Science Reviews}{195}{}{125–151}.
\newblock
\begin{APACrefDOI} \doi{10.1007/s11214-013-0029-z} \end{APACrefDOI}
\PrintBackRefs{\CurrentBib}

\bibitem [\protect \citeauthoryear {%
Haynes%
, Lagerquist%
, McGraw%
, Musgrave%
\BCBL {}\ \BBA {} Ebert-Uphoff%
}{%
Haynes%
\ \protect \BOthers {.}}{%
{\protect \APACyear {2023}}%
}]{%
Haynes2023}
\APACinsertmetastar {%
Haynes2023}%
\begin{APACrefauthors}%
Haynes, K.%
, Lagerquist, R.%
, McGraw, M.%
, Musgrave, K.%
\BCBL {}\ \BBA {} Ebert-Uphoff, I.%
\end{APACrefauthors}%
\unskip\
\newblock
\APACrefYearMonthDay{2023}{}{}.
\newblock
{\BBOQ}\APACrefatitle {Creating and Evaluating Uncertainty Estimates with
  Neural Networks for Environmental Science Applications} {Creating and
  evaluating uncertainty estimates with neural networks for environmental
  science applications}.{\BBCQ}
\newblock
\APACjournalVolNumPages{Artificial Intelligence for the Earth
  Systems}{2}{2}{220061}.
\newblock
\begin{APACrefDOI} \doi{10.1175/AIES-D-22-0061.1} \end{APACrefDOI}
\PrintBackRefs{\CurrentBib}

\bibitem [\protect \citeauthoryear {%
He%
\ \protect \BOthers {.}}{%
He%
\ \protect \BOthers {.}}{%
{\protect \APACyear {2023}}%
}]{%
He2023}
\APACinsertmetastar {%
He2023}%
\begin{APACrefauthors}%
He, F.%
, Fan, K.%
, Hughes, A.%
, Wei, Y.%
, Cui, J.%
, Schneider, N.%
\BDBL {}Zhang, X\BHBI X.%
\end{APACrefauthors}%
\unskip\
\newblock
\APACrefYearMonthDay{2023}{}{}.
\newblock
{\BBOQ}\APACrefatitle {Martian Proton Aurora Brightening Reveals Atmospheric
  Ion Loss Intensifying} {Martian proton aurora brightening reveals atmospheric
  ion loss intensifying}.{\BBCQ}
\newblock
\APACjournalVolNumPages{Geophysical Research Letters}{50}{5}{e2023GL102723}.
\newblock
\APACrefnote{e2023GL102723 2023GL102723}
\newblock
\begin{APACrefDOI} \doi{10.1029/2023GL102723} \end{APACrefDOI}
\PrintBackRefs{\CurrentBib}

\bibitem [\protect \citeauthoryear {%
Hurley%
, Dong%
, Fang%
\BCBL {}\ \BBA {} Brain%
}{%
Hurley%
\ \protect \BOthers {.}}{%
{\protect \APACyear {2018}}%
}]{%
Hurley2018}
\APACinsertmetastar {%
Hurley2018}%
\begin{APACrefauthors}%
Hurley, D\BPBI M.%
, Dong, Y.%
, Fang, X.%
\BCBL {}\ \BBA {} Brain, D\BPBI A.%
\end{APACrefauthors}%
\unskip\
\newblock
\APACrefYearMonthDay{2018}{}{}.
\newblock
{\BBOQ}\APACrefatitle {A Proxy for the Upstream {IMF} Clock Angle Using {MAVEN}
  Magnetic Field Data} {A proxy for the upstream {IMF} clock angle using
  {MAVEN} magnetic field data}.{\BBCQ}
\newblock
\APACjournalVolNumPages{Journal of Geophysical Research: Space
  Physics}{123}{11}{9612-9618}.
\newblock
\begin{APACrefDOI} \doi{10.1029/2018JA025578} \end{APACrefDOI}
\PrintBackRefs{\CurrentBib}

\bibitem [\protect \citeauthoryear {%
B.~Jakosky%
\ \protect \BOthers {.}}{%
B.~Jakosky%
\ \protect \BOthers {.}}{%
{\protect \APACyear {2018}}%
}]{%
Jakosky2018}
\APACinsertmetastar {%
Jakosky2018}%
\begin{APACrefauthors}%
Jakosky, B.%
, Brain, D.%
, Chaffin, M.%
, Curry, S.%
, Deighan, J.%
, Grebowsky, J.%
\BDBL {}Zurek, R.%
\end{APACrefauthors}%
\unskip\
\newblock
\APACrefYearMonthDay{2018}{}{}.
\newblock
{\BBOQ}\APACrefatitle {Loss of the {Martian} atmosphere to space: {P}resent-day
  loss rates determined from {MAVEN} observations and integrated loss through
  time} {Loss of the {Martian} atmosphere to space: {P}resent-day loss rates
  determined from {MAVEN} observations and integrated loss through
  time}.{\BBCQ}
\newblock
\APACjournalVolNumPages{Icarus}{315}{}{146-157}.
\newblock
\begin{APACrefDOI} \doi{10.1016/j.icarus.2018.05.030} \end{APACrefDOI}
\PrintBackRefs{\CurrentBib}

\bibitem [\protect \citeauthoryear {%
B\BPBI M.~Jakosky%
, Grebowsky%
, Luhmann%
\BCBL {}\ \BBA {} Brain%
}{%
B\BPBI M.~Jakosky%
\ \protect \BOthers {.}}{%
{\protect \APACyear {2015}}%
}]{%
Jakosky2015}
\APACinsertmetastar {%
Jakosky2015}%
\begin{APACrefauthors}%
Jakosky, B\BPBI M.%
, Grebowsky, J\BPBI M.%
, Luhmann, J\BPBI G.%
\BCBL {}\ \BBA {} Brain, D\BPBI A.%
\end{APACrefauthors}%
\unskip\
\newblock
\APACrefYearMonthDay{2015}{}{}.
\newblock
{\BBOQ}\APACrefatitle {Initial results from the {MAVEN} mission to {Mars}}
  {Initial results from the {MAVEN} mission to {Mars}}.{\BBCQ}
\newblock
\APACjournalVolNumPages{Geophysical Research Letters}{42}{21}{8791-8802}.
\newblock
\begin{APACrefDOI} \doi{10.1002/2015GL065271} \end{APACrefDOI}
\PrintBackRefs{\CurrentBib}

\bibitem [\protect \citeauthoryear {%
Karpatne%
, Ebert-Uphoff%
, Ravela%
, Babaie%
\BCBL {}\ \BBA {} Kumar%
}{%
Karpatne%
\ \protect \BOthers {.}}{%
{\protect \APACyear {2019}}%
}]{%
Karpatne2019}
\APACinsertmetastar {%
Karpatne2019}%
\begin{APACrefauthors}%
Karpatne, A.%
, Ebert-Uphoff, I.%
, Ravela, S.%
, Babaie, H\BPBI A.%
\BCBL {}\ \BBA {} Kumar, V.%
\end{APACrefauthors}%
\unskip\
\newblock
\APACrefYearMonthDay{2019}{}{}.
\newblock
{\BBOQ}\APACrefatitle {{Machine Learning for the Geosciences: Challenges and
  Opportunities}} {{Machine Learning for the Geosciences: Challenges and
  Opportunities}}.{\BBCQ}
\newblock
\APACjournalVolNumPages{IEEE Transactions on Knowledge and Data
  Engineering}{31}{8}{1544--1554}.
\newblock
\begin{APACrefDOI} \doi{10.1109/TKDE.2018.2861006} \end{APACrefDOI}
\PrintBackRefs{\CurrentBib}

\bibitem [\protect \citeauthoryear {%
Keebler%
, Tóth%
, Zieger%
\BCBL {}\ \BBA {} Opher%
}{%
Keebler%
\ \protect \BOthers {.}}{%
{\protect \APACyear {2022}}%
}]{%
Keebler2022}
\APACinsertmetastar {%
Keebler2022}%
\begin{APACrefauthors}%
Keebler, T\BPBI B.%
, Tóth, G.%
, Zieger, B.%
\BCBL {}\ \BBA {} Opher, M.%
\end{APACrefauthors}%
\unskip\
\newblock
\APACrefYearMonthDay{2022}{}{}.
\newblock
{\BBOQ}\APACrefatitle {MSWIM2D: Two-dimensional Outer Heliosphere Solar Wind
  Modeling} {Mswim2d: Two-dimensional outer heliosphere solar wind
  modeling}.{\BBCQ}
\newblock
\APACjournalVolNumPages{The Astrophysical Journal Supplement
  Series}{260}{2}{43}.
\newblock
\begin{APACrefDOI} \doi{10.3847/1538-4365/ac67eb} \end{APACrefDOI}
\PrintBackRefs{\CurrentBib}

\bibitem [\protect \citeauthoryear {%
King%
\ \BBA {} Papitashvili%
}{%
King%
\ \BBA {} Papitashvili%
}{%
{\protect \APACyear {2005}}%
}]{%
King2005}
\APACinsertmetastar {%
King2005}%
\begin{APACrefauthors}%
King, J\BPBI H.%
\BCBT {}\ \BBA {} Papitashvili, N\BPBI E.%
\end{APACrefauthors}%
\unskip\
\newblock
\APACrefYearMonthDay{2005}{}{}.
\newblock
{\BBOQ}\APACrefatitle {Solar wind spatial scales in and comparisons of hourly
  {Wind} and {ACE} plasma and magnetic field data} {Solar wind spatial scales
  in and comparisons of hourly {Wind} and {ACE} plasma and magnetic field
  data}.{\BBCQ}
\newblock
\APACjournalVolNumPages{Journal of Geophysical Research: Space
  Physics}{110}{A2}{}.
\newblock
\begin{APACrefDOI} \doi{10.1029/2004JA010649} \end{APACrefDOI}
\PrintBackRefs{\CurrentBib}

\bibitem [\protect \citeauthoryear {%
Langlais%
, Thébault%
, Houliez%
, Purucker%
\BCBL {}\ \BBA {} Lillis%
}{%
Langlais%
\ \protect \BOthers {.}}{%
{\protect \APACyear {2019}}%
}]{%
Langlais2019}
\APACinsertmetastar {%
Langlais2019}%
\begin{APACrefauthors}%
Langlais, B.%
, Thébault, E.%
, Houliez, A.%
, Purucker, M\BPBI E.%
\BCBL {}\ \BBA {} Lillis, R\BPBI J.%
\end{APACrefauthors}%
\unskip\
\newblock
\APACrefYearMonthDay{2019}{}{}.
\newblock
{\BBOQ}\APACrefatitle {A New Model of the Crustal Magnetic Field of {Mars}
  Using {MGS} and {MAVEN}} {A new model of the crustal magnetic field of {Mars}
  using {MGS} and {MAVEN}}.{\BBCQ}
\newblock
\APACjournalVolNumPages{Journal of Geophysical Research:
  Planets}{124}{6}{1542-1569}.
\newblock
\begin{APACrefDOI} \doi{10.1029/2018JE005854} \end{APACrefDOI}
\PrintBackRefs{\CurrentBib}

\bibitem [\protect \citeauthoryear {%
Lee%
\ \protect \BOthers {.}}{%
Lee%
\ \protect \BOthers {.}}{%
{\protect \APACyear {2017}}%
}]{%
Lee2017}
\APACinsertmetastar {%
Lee2017}%
\begin{APACrefauthors}%
Lee, C\BPBI O.%
, Hara, T.%
, Halekas, J\BPBI S.%
, Thiemann, E.%
, Chamberlin, P.%
, Eparvier, F.%
\BDBL {}Jakosky, B\BPBI M.%
\end{APACrefauthors}%
\unskip\
\newblock
\APACrefYearMonthDay{2017}{}{}.
\newblock
{\BBOQ}\APACrefatitle {MAVEN observations of the solar cycle 24 space weather
  conditions at Mars} {Maven observations of the solar cycle 24 space weather
  conditions at mars}.{\BBCQ}
\newblock
\APACjournalVolNumPages{Journal of Geophysical Research: Space
  Physics}{122}{3}{2768-2794}.
\newblock
\begin{APACrefDOI} \doi{10.1002/2016JA023495} \end{APACrefDOI}
\PrintBackRefs{\CurrentBib}

\bibitem [\protect \citeauthoryear {%
Lee%
\ \protect \BOthers {.}}{%
Lee%
\ \protect \BOthers {.}}{%
{\protect \APACyear {2018}}%
}]{%
Lee2018}
\APACinsertmetastar {%
Lee2018}%
\begin{APACrefauthors}%
Lee, C\BPBI O.%
, Jakosky, B\BPBI M.%
, Luhmann, J\BPBI G.%
, Brain, D\BPBI A.%
, Mays, M\BPBI L.%
, Hassler, D\BPBI M.%
\BDBL {}Halekas, J\BPBI S.%
\end{APACrefauthors}%
\unskip\
\newblock
\APACrefYearMonthDay{2018}{}{}.
\newblock
{\BBOQ}\APACrefatitle {Observations and Impacts of the 10 {September} 2017
  Solar Events at {Mars}: An Overview and Synthesis of the Initial Results}
  {Observations and impacts of the 10 {September} 2017 solar events at {Mars}:
  An overview and synthesis of the initial results}.{\BBCQ}
\newblock
\APACjournalVolNumPages{Geophysical Research Letters}{45}{17}{8871-8885}.
\newblock
\begin{APACrefDOI} \doi{10.1029/2018GL079162} \end{APACrefDOI}
\PrintBackRefs{\CurrentBib}

\bibitem [\protect \citeauthoryear {%
Lee%
\ \protect \BOthers {.}}{%
Lee%
\ \protect \BOthers {.}}{%
{\protect \APACyear {2023}}%
}]{%
Lee2023}
\APACinsertmetastar {%
Lee2023}%
\begin{APACrefauthors}%
Lee, C\BPBI O.%
, Sánchez-Cano, B.%
, DiBraccio, G\BPBI A.%
, Mayyasi, M.%
, Xu, S.%
, Chamberlin, P.%
\BDBL {}Elliott, H.%
\end{APACrefauthors}%
\unskip\
\newblock
\APACrefYearMonthDay{2023}{}{}.
\newblock
{\BBOQ}\APACrefatitle {Heliophysics and space weather science at $\sim$1.5
  {AU}: {K}nowledge gaps and need for space weather monitors at {Mars}}
  {Heliophysics and space weather science at $\sim$1.5 {AU}: {K}nowledge gaps
  and need for space weather monitors at {Mars}}.{\BBCQ}
\newblock
\APACjournalVolNumPages{Frontiers in Astronomy and Space Sciences}{10}{}{}.
\newblock
\begin{APACrefDOI} \doi{10.3389/fspas.2023.1064208} \end{APACrefDOI}
\PrintBackRefs{\CurrentBib}

\bibitem [\protect \citeauthoryear {%
Liemohn%
\ \protect \BOthers {.}}{%
Liemohn%
\ \protect \BOthers {.}}{%
{\protect \APACyear {2021}}%
}]{%
Liemohn2021}
\APACinsertmetastar {%
Liemohn2021}%
\begin{APACrefauthors}%
Liemohn, M\BPBI W.%
, Shane, A\BPBI D.%
, Azari, A\BPBI R.%
, Petersen, A\BPBI K.%
, Swiger, B\BPBI M.%
\BCBL {}\ \BBA {} Mukhopadhyay, A.%
\end{APACrefauthors}%
\unskip\
\newblock
\APACrefYearMonthDay{2021}{}{}.
\newblock
{\BBOQ}\APACrefatitle {{RMSE} is not enough: {Guidelines} to robust data-model
  comparisons for magnetospheric physics} {{RMSE} is not enough: {Guidelines}
  to robust data-model comparisons for magnetospheric physics}.{\BBCQ}
\newblock
\APACjournalVolNumPages{Journal of Atmospheric and Solar-Terrestrial
  Physics}{218}{}{105624}.
\newblock
\begin{APACrefDOI} \doi{10.1016/j.jastp.2021.105624} \end{APACrefDOI}
\PrintBackRefs{\CurrentBib}

\bibitem [\protect \citeauthoryear {%
{Lillis}%
\ \protect \BOthers {.}}{%
{Lillis}%
\ \protect \BOthers {.}}{%
{\protect \APACyear {2022}}%
}]{%
Lillis2022}
\APACinsertmetastar {%
Lillis2022}%
\begin{APACrefauthors}%
{Lillis}, R\BPBI J.%
, {Curry}, S\BPBI M.%
, {Ma}, Y\BPBI J.%
, {Curtis}, D\BPBI W.%
, {Taylor}, E\BPBI R.%
, {Parker}, J\BPBI S.%
\BDBL {}{Mandy}, C.%
\end{APACrefauthors}%
\unskip\
\newblock
\APACrefYearMonthDay{2022}{}{}.
\newblock
{\BBOQ}\APACrefatitle {{ESCAPADE}: A Twin-Spacecraft {SIMPLEX} Mission to
  Unveil {M}ars' Unique Hybrid Magnetosphere} {{ESCAPADE}: A twin-spacecraft
  {SIMPLEX} mission to unveil {M}ars' unique hybrid magnetosphere}.{\BBCQ}
\newblock
\BIn{} \APACrefbtitle {Low-Cost Science Mission Concepts for {M}ars
  Exploration} {Low-cost science mission concepts for {M}ars exploration}\
  (\BVOL\ 2655, \BPG~5012).
\PrintBackRefs{\CurrentBib}

\bibitem [\protect \citeauthoryear {%
Liu%
\ \protect \BOthers {.}}{%
Liu%
\ \protect \BOthers {.}}{%
{\protect \APACyear {2021}}%
}]{%
Liu2021}
\APACinsertmetastar {%
Liu2021}%
\begin{APACrefauthors}%
Liu, D.%
, Rong, Z.%
, Gao, J.%
, He, J.%
, Klinger, L.%
, Dunlop, M\BPBI W.%
\BDBL {}Wei, Y.%
\end{APACrefauthors}%
\unskip\
\newblock
\APACrefYearMonthDay{2021}{}{}.
\newblock
{\BBOQ}\APACrefatitle {Statistical Properties of Solar Wind Upstream of Mars:
  MAVEN Observations} {Statistical properties of solar wind upstream of mars:
  Maven observations}.{\BBCQ}
\newblock
\APACjournalVolNumPages{The Astrophysical Journal}{911}{2}{113}.
\newblock
\begin{APACrefDOI} \doi{10.3847/1538-4357/abed50} \end{APACrefDOI}
\PrintBackRefs{\CurrentBib}

\bibitem [\protect \citeauthoryear {%
Mackay%
}{%
Mackay%
}{%
{\protect \APACyear {1988}}%
}]{%
Mackay1988}
\APACinsertmetastar {%
Mackay1988}%
\begin{APACrefauthors}%
Mackay, D.%
\end{APACrefauthors}%
\unskip\
\newblock
\APACrefYear{1988}.
\newblock
\APACrefbtitle {Neural Networks and Machine Learning} {Neural networks and
  machine learning}\ (C.~Bishop, \BED{}).
\newblock
\APACaddressPublisher{}{Springer}.
\PrintBackRefs{\CurrentBib}

\bibitem [\protect \citeauthoryear {%
Matthews%
\ \protect \BOthers {.}}{%
Matthews%
\ \protect \BOthers {.}}{%
{\protect \APACyear {2017}}%
}]{%
Matthews2017}
\APACinsertmetastar {%
Matthews2017}%
\begin{APACrefauthors}%
Matthews, A\BPBI G\BPBI d\BPBI G.%
, {van der Wilk}, M.%
, Nickson, T.%
, Fujii, K.%
, {Boukouvalas}, A.%
, {Le{\'o}n-Villagr{\'a}}, P.%
\BDBL {}Hensman, J.%
\end{APACrefauthors}%
\unskip\
\newblock
\APACrefYearMonthDay{2017}{}{}.
\newblock
{\BBOQ}\APACrefatitle {{ {GP}flow: A {G}aussian process library using
  {T}ensor{F}low}} {{ {GP}flow: A {G}aussian process library using
  {T}ensor{F}low}}.{\BBCQ}
\newblock
\APACjournalVolNumPages{Journal of Machine Learning Research}{18}{40}{1-6}.
\PrintBackRefs{\CurrentBib}

\bibitem [\protect \citeauthoryear {%
Mitchell%
\ \protect \BOthers {.}}{%
Mitchell%
\ \protect \BOthers {.}}{%
{\protect \APACyear {2016}}%
}]{%
Mitchell2016}
\APACinsertmetastar {%
Mitchell2016}%
\begin{APACrefauthors}%
Mitchell, D\BPBI L.%
, Mazelle, C.%
, Sauvaud, J\BHBI A.%
, Thocaven, J\BHBI J.%
, Rouzaud, J.%
, Fedorov, A.%
\BDBL {}Jakosky, B\BPBI M.%
\end{APACrefauthors}%
\unskip\
\newblock
\APACrefYearMonthDay{2016}{}{}.
\newblock
{\BBOQ}\APACrefatitle {The {MAVEN Solar Wind Electron Analyzer}} {The {MAVEN
  Solar Wind Electron Analyzer}}.{\BBCQ}
\newblock
\APACjournalVolNumPages{Space Science Reviews}{200}{}{495-528}.
\newblock
\begin{APACrefDOI} \doi{10.1007/s11214-015-0232-1} \end{APACrefDOI}
\PrintBackRefs{\CurrentBib}

\bibitem [\protect \citeauthoryear {%
Morales%
\ \BBA {} Nocedal%
}{%
Morales%
\ \BBA {} Nocedal%
}{%
{\protect \APACyear {2011}}%
}]{%
Morales2011}
\APACinsertmetastar {%
Morales2011}%
\begin{APACrefauthors}%
Morales, J\BPBI L.%
\BCBT {}\ \BBA {} Nocedal, J.%
\end{APACrefauthors}%
\unskip\
\newblock
\APACrefYearMonthDay{2011}{dec}{}.
\newblock
{\BBOQ}\APACrefatitle {Remark on “Algorithm 778: L-BFGS-B: Fortran
  Subroutines for Large-Scale Bound Constrained Optimization”} {Remark on
  “algorithm 778: L-bfgs-b: Fortran subroutines for large-scale bound
  constrained optimization”}.{\BBCQ}
\newblock
\APACjournalVolNumPages{ACM Trans. Math. Softw.}{38}{1}{}.
\newblock
\begin{APACrefDOI} \doi{10.1145/2049662.2049669} \end{APACrefDOI}
\PrintBackRefs{\CurrentBib}

\bibitem [\protect \citeauthoryear {%
{NASA PDS}%
}{%
{NASA PDS}%
}{%
{\protect \APACyear {2023}}%
}]{%
PDSMag}
\APACinsertmetastar {%
PDSMag}%
\begin{APACrefauthors}%
{NASA PDS}.%
\end{APACrefauthors}%
\unskip\
\newblock
\APACrefYearMonthDay{2023}{}{}.
\newblock
\APACrefbtitle {{MAVEN MAG} Calibrated Data Bundle} {{MAVEN MAG} calibrated
  data bundle}\ [dataset].
\newblock
\APACaddressPublisher{}{NASA Planetary Data System}.
\newblock
\begin{APACrefURL}
  \url{https://pds-ppi.igpp.ucla.edu/search/view/?id=pds://PPI/maven.mag.calibrated}
  \end{APACrefURL}
\newblock
\begin{APACrefDOI} \doi{10.17189/1414178} \end{APACrefDOI}
\PrintBackRefs{\CurrentBib}

\bibitem [\protect \citeauthoryear {%
Pedregosa%
\ \protect \BOthers {.}}{%
Pedregosa%
\ \protect \BOthers {.}}{%
{\protect \APACyear {2011}}%
}]{%
scikit-learn}
\APACinsertmetastar {%
scikit-learn}%
\begin{APACrefauthors}%
Pedregosa, F.%
, Varoquaux, G.%
, Gramfort, A.%
, Michel, V.%
, Thirion, B.%
, Grisel, O.%
\BDBL {}Duchesnay, E.%
\end{APACrefauthors}%
\unskip\
\newblock
\APACrefYearMonthDay{2011}{}{}.
\newblock
{\BBOQ}\APACrefatitle {Scikit-learn: Machine Learning in {P}ython}
  {Scikit-learn: Machine learning in {P}ython}.{\BBCQ}
\newblock
\APACjournalVolNumPages{Journal of Machine Learning Research}{12}{}{2825-2830}.
\PrintBackRefs{\CurrentBib}

\bibitem [\protect \citeauthoryear {%
{Plotly Technologies Inc.}%
}{%
{Plotly Technologies Inc.}%
}{%
{\protect \APACyear {2015}}%
}]{%
plotly}
\APACinsertmetastar {%
plotly}%
\begin{APACrefauthors}%
{Plotly Technologies Inc.}%
\end{APACrefauthors}%
\unskip\
\newblock
\APACrefYearMonthDay{2015}{}{}.
\newblock
\APACrefbtitle {Collaborative data science.} {Collaborative data science.}
\newblock
\APACaddressPublisher{Montreal, QC}{Plotly Technologies Inc.}
\newblock
\begin{APACrefURL} \url{https://plot.ly} \end{APACrefURL}
\PrintBackRefs{\CurrentBib}

\bibitem [\protect \citeauthoryear {%
Poduval%
\ \protect \BOthers {.}}{%
Poduval%
\ \protect \BOthers {.}}{%
{\protect \APACyear {2023}}%
}]{%
Poduval2023}
\APACinsertmetastar {%
Poduval2023}%
\begin{APACrefauthors}%
Poduval, B.%
, McPherron, R\BPBI L.%
, Walker, R.%
, Himes, M\BPBI D.%
, Pitman, K\BPBI M.%
, Azari, A\BPBI R.%
\BDBL {}Wing, S.%
\end{APACrefauthors}%
\unskip\
\newblock
\APACrefYearMonthDay{2023}{}{}.
\newblock
{\BBOQ}\APACrefatitle {AI-ready data in space science and solar physics:
  problems, mitigation and action plan} {Ai-ready data in space science and
  solar physics: problems, mitigation and action plan}.{\BBCQ}
\newblock
\APACjournalVolNumPages{Frontiers in Astronomy and Space Sciences}{10}{}{}.
\newblock
\begin{APACrefDOI} \doi{10.3389/fspas.2023.1203598} \end{APACrefDOI}
\PrintBackRefs{\CurrentBib}

\bibitem [\protect \citeauthoryear {%
Pérez%
\ \protect \BOthers {.}}{%
Pérez%
\ \protect \BOthers {.}}{%
{\protect \APACyear {2019}}%
}]{%
Pérez2019}
\APACinsertmetastar {%
Pérez2019}%
\begin{APACrefauthors}%
Pérez, F.%
, Hamman, J.%
, Larsen, L.%
, Paul, K.%
, Heagy, L.%
, Holdgraf, C.%
\BCBL {}\ \BBA {} Panda, Y.%
\end{APACrefauthors}%
\unskip\
\newblock
\APACrefYearMonthDay{2019}{}{}.
\newblock
{\BBOQ}\APACrefatitle {Jupyter meets the Earth: Enabling discovery in
  geoscience through interactive computing at scale} {Jupyter meets the earth:
  Enabling discovery in geoscience through interactive computing at
  scale}.{\BBCQ}
\newblock
\APACjournalVolNumPages{Zenodo}{}{}{}.
\newblock
\begin{APACrefDOI} \doi{10.5281/zenodo.3369939} \end{APACrefDOI}
\PrintBackRefs{\CurrentBib}

\bibitem [\protect \citeauthoryear {%
Qui\~{n}onero Candela%
\ \BBA {} Rasmussen%
}{%
Qui\~{n}onero Candela%
\ \BBA {} Rasmussen%
}{%
{\protect \APACyear {2005}}%
}]{%
Candela2005}
\APACinsertmetastar {%
Candela2005}%
\begin{APACrefauthors}%
Qui\~{n}onero Candela, J.%
\BCBT {}\ \BBA {} Rasmussen, C\BPBI E.%
\end{APACrefauthors}%
\unskip\
\newblock
\APACrefYearMonthDay{2005}{dec}{}.
\newblock
{\BBOQ}\APACrefatitle {A Unifying View of Sparse Approximate Gaussian Process
  Regression} {A unifying view of sparse approximate gaussian process
  regression}.{\BBCQ}
\newblock
\APACjournalVolNumPages{J. Mach. Learn. Res.}{6}{}{1939–1959}.
\PrintBackRefs{\CurrentBib}

\bibitem [\protect \citeauthoryear {%
Ramstad%
\ \protect \BOthers {.}}{%
Ramstad%
\ \protect \BOthers {.}}{%
{\protect \APACyear {2023}}%
}]{%
Ramstad2023}
\APACinsertmetastar {%
Ramstad2023}%
\begin{APACrefauthors}%
Ramstad, R.%
, Brain, D\BPBI A.%
, Dong, Y.%
, Halekas, J\BPBI S.%
, McFadden, J\BPBI M.%
, Mitchell, D\BPBI L.%
\BDBL {}Jakosky, B\BPBI M.%
\end{APACrefauthors}%
\unskip\
\newblock
\APACrefYearMonthDay{2023}{}{}.
\newblock
{\BBOQ}\APACrefatitle {Solar wind driven influences on the {M}artian oxygen
  corona: Constraints on atmospheric sputtering from a synthesis of {MAVEN}
  measurements during solar minimum} {Solar wind driven influences on the
  {M}artian oxygen corona: Constraints on atmospheric sputtering from a
  synthesis of {MAVEN} measurements during solar minimum}.{\BBCQ}
\newblock
\APACjournalVolNumPages{Icarus}{397}{}{115491}.
\newblock
\begin{APACrefDOI} \doi{10.1016/j.icarus.2023.115491} \end{APACrefDOI}
\PrintBackRefs{\CurrentBib}

\bibitem [\protect \citeauthoryear {%
Ramstad%
, Holmstrom%
, Barabash%
\BCBL {}\ \BBA {} Lundin%
}{%
Ramstad%
\ \protect \BOthers {.}}{%
{\protect \APACyear {2017}}%
}]{%
Ramstad2017}
\APACinsertmetastar {%
Ramstad2017}%
\begin{APACrefauthors}%
Ramstad, R.%
, Holmstrom, M.%
, Barabash, S.%
\BCBL {}\ \BBA {} Lundin, R.%
\end{APACrefauthors}%
\unskip\
\newblock
\APACrefYearMonthDay{2017}{}{}.
\newblock
{\BBOQ}\APACrefatitle {{MARS EXPRESS ASPERA-3 IMA SOLAR WIND PARAMETERS V1.0,
  MEX-SUN-ASPERA3-4-SWM-V1.0}} {{MARS EXPRESS ASPERA-3 IMA SOLAR WIND
  PARAMETERS V1.0, MEX-SUN-ASPERA3-4-SWM-V1.0}}.{\BBCQ}
\newblock
\APACjournalVolNumPages{ESA Planetary Science Archive (PSA)}{}{}{}.
\newblock
\begin{APACrefURL}
  \url{https://pds.nasa.gov/ds-view/pds/viewProfile.jsp?dsid=MEX-SUN-ASPERA3-4-SWM-V1.0}
  \end{APACrefURL}
\PrintBackRefs{\CurrentBib}

\bibitem [\protect \citeauthoryear {%
Rasmussen%
\ \BBA {} Williams%
}{%
Rasmussen%
\ \BBA {} Williams%
}{%
{\protect \APACyear {2006}}%
}]{%
Rasmussen2006}
\APACinsertmetastar {%
Rasmussen2006}%
\begin{APACrefauthors}%
Rasmussen, C\BPBI E.%
\BCBT {}\ \BBA {} Williams, C\BPBI K\BPBI I.%
\end{APACrefauthors}%
\unskip\
\newblock
\APACrefYear{2006}.
\newblock
\APACrefbtitle {Gaussian Processes for Machine Learning} {Gaussian processes
  for machine learning}.
\newblock
\APACaddressPublisher{}{the MIT Press}.
\PrintBackRefs{\CurrentBib}

\bibitem [\protect \citeauthoryear {%
Rudin%
}{%
Rudin%
}{%
{\protect \APACyear {2019}}%
}]{%
Rudin2019}
\APACinsertmetastar {%
Rudin2019}%
\begin{APACrefauthors}%
Rudin, C.%
\end{APACrefauthors}%
\unskip\
\newblock
\APACrefYearMonthDay{2019}{}{}.
\newblock
{\BBOQ}\APACrefatitle {{Stop explaining black box machine learning models for
  high stakes decisions and use interpretable models instead}} {{Stop
  explaining black box machine learning models for high stakes decisions and
  use interpretable models instead}}.{\BBCQ}
\newblock
\APACjournalVolNumPages{Nature Machine Intelligence}{1}{}{206–215}.
\newblock
\begin{APACrefDOI} \doi{10.1038/s42256-019-0048-x} \end{APACrefDOI}
\PrintBackRefs{\CurrentBib}

\bibitem [\protect \citeauthoryear {%
Ruhunusiri%
\ \protect \BOthers {.}}{%
Ruhunusiri%
\ \protect \BOthers {.}}{%
{\protect \APACyear {2018}}%
}]{%
Ruhunusiri2018}
\APACinsertmetastar {%
Ruhunusiri2018}%
\begin{APACrefauthors}%
Ruhunusiri, S.%
, Halekas, J\BPBI S.%
, Espley, J\BPBI R.%
, Eparvier, F.%
, Brain, D.%
, Mazelle, C.%
\BDBL {}Jakosky, B\BPBI M.%
\end{APACrefauthors}%
\unskip\
\newblock
\APACrefYearMonthDay{2018}{}{}.
\newblock
{\BBOQ}\APACrefatitle {An Artificial Neural Network for Inferring Solar Wind
  Proxies at {Mars}} {An artificial neural network for inferring solar wind
  proxies at {Mars}}.{\BBCQ}
\newblock
\APACjournalVolNumPages{Geophysical Research Letters}{45}{20}{10,855-10,865}.
\newblock
\begin{APACrefDOI} \doi{10.1029/2018GL079282} \end{APACrefDOI}
\PrintBackRefs{\CurrentBib}

\bibitem [\protect \citeauthoryear {%
{Sanchez-Cano}%
, {Opgenoorth}%
, {Leblanc}%
, {Andrews}%
\BCBL {}\ \BBA {} {Lester}%
}{%
{Sanchez-Cano}%
\ \protect \BOthers {.}}{%
{\protect \APACyear {2022}}%
}]{%
Sanchez-Cano2022}
\APACinsertmetastar {%
Sanchez-Cano2022}%
\begin{APACrefauthors}%
{Sanchez-Cano}, B.%
, {Opgenoorth}, H.%
, {Leblanc}, F.%
, {Andrews}, D.%
\BCBL {}\ \BBA {} {Lester}, M.%
\end{APACrefauthors}%
\unskip\
\newblock
\APACrefYearMonthDay{2022}{}{}.
\newblock
{\BBOQ}\APACrefatitle {{The M-MATISSE mission: Mars Magnetosphere ATmosphere
  Ionosphere and Surface SciencE}} {{The M-MATISSE mission: Mars Magnetosphere
  ATmosphere Ionosphere and Surface SciencE}}.{\BBCQ}
\newblock
\BIn{} \APACrefbtitle {44th COSPAR Scientific Assembly} {44th cospar scientific
  assembly}\ (\BVOL~44, \BPG~421).
\PrintBackRefs{\CurrentBib}

\bibitem [\protect \citeauthoryear {%
Schneider%
\ \protect \BOthers {.}}{%
Schneider%
\ \protect \BOthers {.}}{%
{\protect \APACyear {2015}}%
}]{%
Schneider2015}
\APACinsertmetastar {%
Schneider2015}%
\begin{APACrefauthors}%
Schneider, N\BPBI M.%
, Deighan, J\BPBI I.%
, Jain, S\BPBI K.%
, Stiepen, A\BPBI I., A.~Stewart%
, Larson, D.%
, Mitchell, D\BPBI L.%
\BDBL {}Jakosky, B\BPBI M.%
\end{APACrefauthors}%
\unskip\
\newblock
\APACrefYearMonthDay{2015}{}{}.
\newblock
{\BBOQ}\APACrefatitle {Discovery of diffuse aurora on {Mars}} {Discovery of
  diffuse aurora on {Mars}}.{\BBCQ}
\newblock
\APACjournalVolNumPages{Science}{350}{}{}.
\newblock
\begin{APACrefDOI} \doi{10.1126/science.aad0313} \end{APACrefDOI}
\PrintBackRefs{\CurrentBib}

\bibitem [\protect \citeauthoryear {%
Swiger%
, Liemohn%
\BCBL {}\ \BBA {} Ganushkina%
}{%
Swiger%
\ \protect \BOthers {.}}{%
{\protect \APACyear {2020}}%
}]{%
Swiger2020}
\APACinsertmetastar {%
Swiger2020}%
\begin{APACrefauthors}%
Swiger, B\BPBI M.%
, Liemohn, M\BPBI W.%
\BCBL {}\ \BBA {} Ganushkina, N\BPBI Y.%
\end{APACrefauthors}%
\unskip\
\newblock
\APACrefYearMonthDay{2020}{}{}.
\newblock
{\BBOQ}\APACrefatitle {Improvement of Plasma Sheet Neural Network Accuracy With
  Inclusion of Physical Information} {Improvement of plasma sheet neural
  network accuracy with inclusion of physical information}.{\BBCQ}
\newblock
\APACjournalVolNumPages{Frontiers in Astronomy and Space Sciences}{7}{}{}.
\newblock
\begin{APACrefDOI} \doi{10.3389/fspas.2020.00042} \end{APACrefDOI}
\PrintBackRefs{\CurrentBib}

\bibitem [\protect \citeauthoryear {%
Tazi%
\ \protect \BOthers {.}}{%
Tazi%
\ \protect \BOthers {.}}{%
{\protect \APACyear {2023}}%
}]{%
Tazi2023}
\APACinsertmetastar {%
Tazi2023}%
\begin{APACrefauthors}%
Tazi, K.%
, Lin, J\BPBI A.%
, Viljoen, R.%
, Gardner, A.%
, John, S\BPBI T.%
, Ge, H.%
\BCBL {}\ \BBA {} Turner, R\BPBI E.%
\end{APACrefauthors}%
\unskip\
\newblock
\APACrefYearMonthDay{2023}{}{}.
\newblock
{\BBOQ}\APACrefatitle {Beyond Intuition, a Framework for Applying {GP}s to
  Real-World Data} {Beyond intuition, a framework for applying {GP}s to
  real-world data}.{\BBCQ}
\newblock
\BIn{} \APACrefbtitle {{ICML} 2023 Workshop on Structured Probabilistic
  Inference {\&} Generative Modeling.} {{ICML} 2023 workshop on structured
  probabilistic inference {\&} generative modeling.}
\newblock
\begin{APACrefURL} \url{https://arxiv.org/pdf/2307.03093.pdf} \end{APACrefURL}
\PrintBackRefs{\CurrentBib}

\bibitem [\protect \citeauthoryear {%
{van der Wilk}%
\ \protect \BOthers {.}}{%
{van der Wilk}%
\ \protect \BOthers {.}}{%
{\protect \APACyear {2020}}%
}]{%
vanderwilk2020}
\APACinsertmetastar {%
vanderwilk2020}%
\begin{APACrefauthors}%
{van der Wilk}, M.%
, Dutordoir, V.%
, John, S.%
, Artemev, A.%
, Adam, V.%
\BCBL {}\ \BBA {} Hensman, J.%
\end{APACrefauthors}%
\unskip\
\newblock
\APACrefYearMonthDay{2020}{}{}.
\newblock
{\BBOQ}\APACrefatitle {A Framework for Interdomain and Multioutput {G}aussian
  Processes} {A framework for interdomain and multioutput {G}aussian
  processes}.{\BBCQ}
\newblock
\APACjournalVolNumPages{arXiv:2003.01115}{}{}{}.
\newblock
\begin{APACrefURL} \url{https://arxiv.org/abs/2003.01115} \end{APACrefURL}
\PrintBackRefs{\CurrentBib}

\bibitem [\protect \citeauthoryear {%
Virtanen%
\ \protect \BOthers {.}}{%
Virtanen%
\ \protect \BOthers {.}}{%
{\protect \APACyear {2020}}%
}]{%
scipy}
\APACinsertmetastar {%
scipy}%
\begin{APACrefauthors}%
Virtanen, P.%
, Gommers, R.%
, Oliphant, T\BPBI E.%
, Haberland, M.%
, Reddy, T.%
, Cournapeau, D.%
\BDBL {}{SciPy 1.0 Contributors}%
\end{APACrefauthors}%
\unskip\
\newblock
\APACrefYearMonthDay{2020}{}{}.
\newblock
{\BBOQ}\APACrefatitle {{{SciPy} 1.0: Fundamental Algorithms for Scientific
  Computing in Python}} {{{SciPy} 1.0: Fundamental Algorithms for Scientific
  Computing in Python}}.{\BBCQ}
\newblock
\APACjournalVolNumPages{Nature Methods}{17}{}{261--272}.
\newblock
\begin{APACrefDOI} \doi{10.1038/s41592-019-0686-2} \end{APACrefDOI}
\PrintBackRefs{\CurrentBib}

\bibitem [\protect \citeauthoryear {%
Wang%
\ \protect \BOthers {.}}{%
Wang%
\ \protect \BOthers {.}}{%
{\protect \APACyear {2023}}%
}]{%
Wang2023}
\APACinsertmetastar {%
Wang2023}%
\begin{APACrefauthors}%
Wang, J.%
, Shi, Y.%
, Luo, B.%
, Liu, S.%
, Kong, L.%
, Ma, J.%
\BDBL {}Chen, Y.%
\end{APACrefauthors}%
\unskip\
\newblock
\APACrefYearMonthDay{2023}{}{}.
\newblock
{\BBOQ}\APACrefatitle {Upstream Solar Wind Prediction up to {Mars} by an
  Operational Solar Wind Prediction System} {Upstream solar wind prediction up
  to {Mars} by an operational solar wind prediction system}.{\BBCQ}
\newblock
\APACjournalVolNumPages{Space Weather}{21}{1}{e2022SW003281}.
\newblock
\APACrefnote{e2022SW003281 2022SW003281}
\newblock
\begin{APACrefDOI} \doi{10.1029/2022SW003281} \end{APACrefDOI}
\PrintBackRefs{\CurrentBib}

\bibitem [\protect \citeauthoryear {%
Zhu%
, Byrd%
, Lu%
\BCBL {}\ \BBA {} Nocedal%
}{%
Zhu%
\ \protect \BOthers {.}}{%
{\protect \APACyear {1997}}%
}]{%
Zhu1997}
\APACinsertmetastar {%
Zhu1997}%
\begin{APACrefauthors}%
Zhu, C.%
, Byrd, R\BPBI H.%
, Lu, P.%
\BCBL {}\ \BBA {} Nocedal, J.%
\end{APACrefauthors}%
\unskip\
\newblock
\APACrefYearMonthDay{1997}{dec}{}.
\newblock
{\BBOQ}\APACrefatitle {Algorithm 778: L-BFGS-B: Fortran Subroutines for
  Large-Scale Bound-Constrained Optimization} {Algorithm 778: L-bfgs-b: Fortran
  subroutines for large-scale bound-constrained optimization}.{\BBCQ}
\newblock
\APACjournalVolNumPages{ACM Trans. Math. Softw.}{23}{4}{550–560}.
\newblock
\begin{APACrefDOI} \doi{10.1145/279232.279236} \end{APACrefDOI}
\PrintBackRefs{\CurrentBib}

\bibitem [\protect \citeauthoryear {%
Zou%
\ \protect \BOthers {.}}{%
Zou%
\ \protect \BOthers {.}}{%
{\protect \APACyear {2021}}%
}]{%
Zou2021}
\APACinsertmetastar {%
Zou2021}%
\begin{APACrefauthors}%
Zou, Y.%
, Zhu, Y.%
, Bai, Y.%
, Wang, L.%
, Jia, Y.%
, Shen, W.%
\BDBL {}Peng, Y.%
\end{APACrefauthors}%
\unskip\
\newblock
\APACrefYearMonthDay{2021}{}{}.
\newblock
{\BBOQ}\APACrefatitle {Scientific objectives and payloads of {T}ianwen-1,
  {C}hina’s first {M}ars exploration mission} {Scientific objectives and
  payloads of {T}ianwen-1, {C}hina’s first {M}ars exploration
  mission}.{\BBCQ}
\newblock
\APACjournalVolNumPages{Advances in Space Research}{67}{2}{812-823}.
\newblock
\begin{APACrefDOI} \doi{10.1016/j.asr.2020.11.005} \end{APACrefDOI}
\PrintBackRefs{\CurrentBib}

\bibitem [\protect \citeauthoryear {%
Zuber%
\ \protect \BOthers {.}}{%
Zuber%
\ \protect \BOthers {.}}{%
{\protect \APACyear {1992}}%
}]{%
Zuber1992}
\APACinsertmetastar {%
Zuber1992}%
\begin{APACrefauthors}%
Zuber, M\BPBI T.%
, Smith, D\BPBI E.%
, Solomon, S\BPBI C.%
, Muhleman, D\BPBI O.%
, Head, J\BPBI W.%
, Garvin, J\BPBI B.%
\BDBL {}Bufton, J\BPBI L.%
\end{APACrefauthors}%
\unskip\
\newblock
\APACrefYearMonthDay{1992}{}{}.
\newblock
{\BBOQ}\APACrefatitle {The Mars Observer laser altimeter investigation} {The
  mars observer laser altimeter investigation}.{\BBCQ}
\newblock
\APACjournalVolNumPages{Journal of Geophysical Research:
  Planets}{97}{E5}{7781-7797}.
\newblock
\begin{APACrefDOI} \doi{10.1029/92JE00341} \end{APACrefDOI}
\PrintBackRefs{\CurrentBib}

\bibitem [\protect \citeauthoryear {%
Zwillinger%
\ \BBA {} Kokoska%
}{%
Zwillinger%
\ \BBA {} Kokoska%
}{%
{\protect \APACyear {2000}}%
}]{%
Zwillinger2000}
\APACinsertmetastar {%
Zwillinger2000}%
\begin{APACrefauthors}%
Zwillinger, D.%
\BCBT {}\ \BBA {} Kokoska, S.%
\end{APACrefauthors}%
\unskip\
\newblock
\APACrefYear{2000}.
\newblock
\APACrefbtitle {CRC Standard Probability and Statistics Tables and Formulae}
  {Crc standard probability and statistics tables and formulae}.
\newblock
\APACaddressPublisher{}{Chapman and Hall: New York}.
\PrintBackRefs{\CurrentBib}

\end{thebibliography}

\end{document}